\documentclass[
    aps,
    prd,
    twocolumn,
    superscriptaddress,
    nofootinbib,
    amsmath,
    amssymb
]{revtex4-2}

\usepackage[T1]{fontenc}          
\usepackage[british]{babel}       
\usepackage{microtype}            

\usepackage{aas_macros}  
\usepackage{graphicx}
\usepackage{dcolumn}
\usepackage{bm}
\usepackage{url}
\usepackage{subcaption}
\usepackage{amsfonts}
\usepackage{array}
\usepackage{multirow}
\usepackage{xcolor}
\usepackage[pdftex, colorlinks=true, allcolors=blue]{hyperref}

\usepackage[ruled,vlined]{algorithm2e}
\usepackage{setspace}
\usepackage{ragged2e}  

\SetAlgoNlRelativeSize{-1}  
\SetInd{0.5em}{1.5em}            
\SetAlgoSkip{smallskip}          
\DontPrintSemicolon               


\makeatletter
\long\def\@makecaption#1#2{%
  \vskip\abovecaptionskip
  \sbox\@tempboxa{\small#1. #2}%
  \ifdim \wd\@tempboxa > \hsize
    {\small\justifying#1. #2\par}
  \else
    \global\@minipagefalse
    \hb@xt@\hsize{\hfil\box\@tempboxa\hfil}%
  \fi
  \vskip\belowcaptionskip}
\makeatother

\newcommand*{\valencia}{Departamento de Astronom\'{\i}a y Astrof\'{\i}sica, Universitat de Val\`encia, Dr.~Moliner 50, 46100 Burjassot (Val\`encia), Spain.}
\newcommand*{\OAUV}{Observatori Astron\`omic, Universitat de Val\`encia, Catedr\'atico Jos\'e Beltr\'an 2, 46980 Paterna (Val\`encia), Spain}

\DeclareMathOperator*{\argmin}{arg\,min}

\begin{document}

\title{\textnormal{\textsc{clawdia}}: A dictionary learning framework for gravitational-wave data analysis}

\author{Miquel Llorens-Monteagudo} \email{Contact author: miquel.llorens@uv.es} \affiliation{\valencia}
\author{Alejandro Torres-Forn\'e} \affiliation{\valencia} \affiliation{\OAUV}
\author{Jos\'e A.~Font} \affiliation{\valencia} \affiliation{\OAUV}

\date{\today}

\begin{abstract}
Deep-learning methods are becoming increasingly important in gravitational-wave data analysis, yet their performance often relies on large training datasets and models whose internal representations are difficult to interpret. Sparse dictionary learning (SDL) offers a complementary approach: it performs well in scarce-data regimes and yields physically interpretable representations of gravitational-wave morphology. Here we present \textsc{clawdia} (Comprehensive Library for the Analysis of Waves via Dictionary-based Algorithms), an open-source Python framework that integrates SDL-based denoising and classification under realistic detector noise. We systematise previously isolated SDL workflows into a unified, modular environment with a consistent, user-friendly interface. The current release provides several time-domain denoising strategies based on LASSO-regularised sparse coding and a classifier based on Low-Rank Shared Dictionary Learning. A companion toolbox, \textsc{gwadama}, supports dataset construction and realistic conditioning of real and simulated interferometer data. We demonstrate \textsc{clawdia}'s performance by denoising the signal from binary neutron star event GW170817 and by classifying families of instrumental glitches from LIGO's third observing run, highlighting robustness at low signal-to-noise ratios. \textsc{clawdia} is intended as a community-driven, interoperable library extensible to additional tasks, including detection and parameter estimation.
\end{abstract}


\maketitle

\section{\label{sec:introduction}INTRODUCTION}

The detection of gravitational waves (GWs) by the Advanced LIGO~\cite{LIGO:2015} and Advanced Virgo~\cite{Acernese:2014hva} observatories has opened the new era of GW astronomy. During their first three observing runs, the LIGO-Virgo-KAGRA (LVK) Collaboration, which operates the network of ground-based detectors  Advanced LIGO, Advanced Virgo, and KAGRA~\cite{KAGRA:2020tym}, reported nearly one hundred compact binary coalescence (CBC) events~\cite{GWTC-1,GWTC-2,GWTC-3}. More recently, in August 2025, the LVK Collaboration has published its fourth Gravitational-Wave Transient Catalog~\cite{GWTC-4}, containing 128 new significant GW candidates, all of them associated with CBCs. The LVK observations include all three classes of CBC systems, namely, binary black holes (BBH), binary neutron stars (BNS), and black hole-neutron star (BHNS) systems. The latter two are prime targets for multi-messenger observations, being visible in both GWs, electromagnetic (EM) radiation, and possibly also accessible to neutrino telescopes. Multi-messenger astronomy with GWs was dramatically realised with BNS merger signal GW170817~\cite{GW170817,Abbott:2017:multi-messenger}, the first GW event with a confirmed EM counterpart~\cite{Coulter:2017}. 

Maximising the scientific return of GW observations requires overcoming significant challenges in signal extraction and data analysis. The vast amount of data generated by current and future detectors, combined with the complexity of extracting weak astrophysical signals from noisy backgrounds, requires the development of robust and efficient analysis techniques~\cite{LVK:2020:noise}. Indeed, the detection of GWs is greatly hindered by numerous noise sources that introduce nonstationary and non-Gaussian artifacts (so-called glitches) into the recorded time-series data~\cite{Nuttall:2018, Zackay:2019}. 

Many astrophysical GW signals, particularly those with low amplitudes, reside at or below the sensitivity limits of current interferometers. Matched filtering continues to be the primary method for identifying well-modeled signals, such as those from CBCs~\cite{LVK:2020:noise}, but its computational cost becomes prohibitive for continuous wave sources like rotating neutron stars~\cite{Keith:2023}. Additionally, unmodeled or poorly modeled transient signals, such as those from core-collapse supernovae (CCSNe), highly eccentric BBH mergers, or hypothetical sources like cosmic string cusps, require alternative data analysis approaches~\cite{Damour:2004, Huerta:2014, Andresen:2019, Powell:2025:DAWES}.

Machine learning (ML) is becoming increasingly ubiquitous in GW data analysis, offering promising alternatives to traditional methods (see~\cite{Benedetto:2023,Stergioulas:2024,Cuoco:2025} for recent reviews). Indeed, ML approaches already permeate many GW data analysis tasks, including glitch classification and denoising, waveform generation through surrogate models, detection searches, and parameter estimation. Neural networks (NNs) have demonstrated remarkable performance in such tasks~\cite{George:2018, Gabbard:2018, Shen:2019, Krastev:2020, Fernandes:2023, Freitas:2024, Koloniari:2025, Freitas:2025, Roberto:2025, Cuoco:2025}, but often suffer from interpretability issues and require large labeled datasets for training. Sparse dictionary learning (SDL)
has emerged as a compelling and robust  alternative, especially in situations where data availability is limited~\cite{Zhang:2015,Tang:2021,Zhang:2022}. SDL achieves the sparse representation of the data through the linear combination of basic elements of the signals making up a dictionary, dubbed as ‘atoms’ in the SDL jargon. This technique has been applied with remarkable success in different scientific fields, demonstrating its usefulness for signal representation~\cite{Zhang:2015:SDLReviewInVC,Yu:2015,Xu:2017,Wang:2021}.
Very recently, the first studies of SDL in the context of GW data analysis have been reported. Those range from the elimination of instrumental noise from the detectors to the reconstruction of signals in different astrophysical settings (see e.g.~\cite{Torres-Forne:2016, Llorens-Monteagudo:2019, Torres-Forne:2020, Saiz-Perez:2022, Badger:2023, Powell:2024,  Badger:2025, Llorens-Monteagudo:2025:bns}). 

In this article, we present \textsc{clawdia}---{\textbf{C}omprehensive \textbf{L}ibrary for the \textbf{A}nalysis of \textbf{W}aves via \textbf{Di}ctionary-based \textbf{A}lgorithms}---an open-source, Python framework for applying SDL methods to GW data analysis. \textsc{clawdia} is designed to formalize and systematize the use of SDL in this context, offering a flexible, modular environment supporting, currently, a variety of workflows for denoising and classification of GW signals. The development of \textsc{clawdia} is motivated by both astrophysical and methodological considerations. On the astrophysical side, SDL techniques offer a compelling alternative to mainstream ML models based on deep NNs, which often face limitations in situations with insufficient training data, low signal-to-noise ratio (SNR), severe class imbalance, or the need for interpretability. Those conditions are frequently encountered when analyzing simulated GW signals as, for example, from CCSNe~\cite{OConnor:2018, Pan:2018, Powell:2019, Radice:2019, Mezzacappa:2020, Powell:2020, Pajkos:2021} or postmerger remnants of BNS mergers~\cite{Oechslin:2007, Bauswein:2012, Bauswein:2015, Topolski:2023}. Such signals are in very short supply for the typical standards of ML datasets. The reasons are because the underlying simulations are intrinsically fairly expensive and, moreover, they show significant morphological variability given the inherently large parameter space of these systems. SDL-based approaches combine robustness to noise, interpretable representations, and reliable performance in data-scarce regimes, making them particularly well suited to handle these challenges. \textsc{clawdia} aims to bring together both existing state-of-the-art SDL techniques and novel developments in a unified framework, enabling users to experiment with different models, parameters, and workflows adapted to a wide variety of astrophysical contexts.

From a technical standpoint, \textsc{clawdia} aims to consolidate current SDL developments in GW data analysis---many of which are currently confined to isolated, publication-specific implementations---into a coherent, community-driven framework. Inspired by foundational scientific software such as \textsc{NumPy}~\cite{Harris:2020:NumPy}, \textsc{SciPy}~\cite{Virtanen:2020:SciPy}, as well as GW-specific projects like \textsc{GWpy}~\cite{Macleod:2021:GWpy} and \textsc{Bilby}~\cite{Ashton:2019:Bilby}, \textsc{clawdia} seeks to establish itself as a standard tool for sparse modeling in GW science. The project emphasizes clarity, extensibility, and ease of use, with proper documentation and ready-to-use examples to foster adoption and reproducibility. Although fully implemented in Python to ensure fast prototyping and usability, future versions of \textsc{clawdia} will progressively integrate compiled components to accelerate performance in data-intensive tasks, while maintaining its accessible Python interface.

At the time of writing, \textsc{clawdia} includes several SDL-based denoising algorithms, a classification model, and a configurable pipeline that allows users to switch between models and adapt workflows to specific problem settings. All current methods operate on time-domain representations of the GW strain, a choice that reduces preprocessing complexity and limits the size of the parameter space to be explored during optimisation. Nevertheless, these methods can be extended to other representations, such as Fourier-domain signals or spectrograms, which may offer advantages in certain scenarios. While its current focus lies in classification, denoising, and reconstruction, \textsc{clawdia} is designed for continuous growth. Upcoming versions will expand its capabilities to explore signal detection, parameter estimation, and potentially even data generation, with the ultimate goal of establishing SDL as a first-class methodology in GW data analysis. The codebase is publicly available on GitHub\footnote{
    \url{https://github.com/MiquelLluis/CLAWDIA}
} and includes API documentation styled after \textsc{NumPy}'s.

The organisation of the paper is as follows: Section~\ref{sec:sparse-dictionary-learning} briefly discusses mathematical and technical aspects of SDL, focusing on the reconstruction, the dictionary-learning problem, and the classification. Section~\ref{sec:CLAWDIA-framework-and-functionality} presents the \textsc{clawdia} framework and its functionality. This section describes the various denoising/reconstruction methods implemented, the corresponding classification approaches, and provides optimisation guidelines. The classification pipeline is discussed in detail in Section~\ref{sec:classification-pipeline} while illustrative results for both denoising and classification are reported in Section~\ref{sec:illustrative-examples}. Finally,  Section~\ref{sec:future-directions} summarizes our work and outlines future extensions. The paper also contains an appendix where we discuss \textsc{gwadama} ({\bf G}ravitational-{\bf Wa}ve {\bf Da}taset {\bf Ma}nager), a Python toolbox developed alongside \textsc{clawdia} to accelerate the management and preparation of datasets for GW data analysis.

\section{SPARSE DICTIONARY LEARNING
\label{sec:sparse-dictionary-learning}}

\subsection{Sparse reconstruction\label{sec:sparse-reconstruction}}

One of the most common challenges in communications and signal processing is the removal of noise, whether in one-dimensional signals or two-dimensional images. Denoising refers to the process of reconstructing a signal removing the noise while preserving the essential structures and details of the original signal or image. This task is a classic example of an inverse problem, where the objective is to recover the original signal from noisy observations. In the context of GW detectors, denoising is crucial to isolate weak GW signals\footnote{
    We use the term \textit{GW signal} to refer to any kind of targeted waveform observable in a GW detector, including non-astrophysical transients such as glitches.
} from background noise, as these signals are often buried under various noise sources, including instrumental and environmental noise.

To solve such inverse problems, we follow the work on GW denoising using dictionaries from Torres-Forn\'e et al~\cite{Torres:2016}, in which the way GWs are embedded into noise is assumed to be described by the linear degradation model
\begin{equation}\label{eq:linear-degradation-model}
\bm{h} = \bm{u} + \bm{n},
\end{equation}
where $\bm{h} \in \mathbb{R}^l$ is the detector strain, and $\bm{u} \in \mathbb{R}^l$ represents the original GW signal. For simplicity, we assume that $\bm{n}$ is Gaussian white noise, meaning it is a square-integrable function with zero mean.

Mallat and Zhang~\cite{Mallat:1993} defined a dictionary as a collection of signals called atoms such that ${\bm{u}}={\bm{D}}\bm{\alpha}$, where ${\bm{u}}$ is the signal to be reconstructed, ${\bm{D}}$ is the dictionary, i.e.~a matrix composed of $a$ atoms of length $l$, and $\bm{\alpha}$ is a sparse $a$-dimensional vector which contains the coefficients of the representation.
Given an overcomplete dictionary $\bm{D} \in \mathbb{R}^{l \times a}$, where the number of atoms $a$ is greater than their length $l$, there is a sparse vector $\bm{\alpha} \in \mathbb{R}^a$ for which $\bm{D}\bm{\alpha} \sim \bm{u}$. Attention must be drawn to the similarity symbol, for the reconstruction ${\bm{D}}\bm{\alpha}$ ought to be closer to the original signal ${\bm{u}}$ than to the strain ${\bm{h}}$. Because the original signal is usually unknown, finding $\bm{\alpha}$ involves imposing a limitation to its similarity to the detector's strain.

Let us consider again the linear degradation model described in Eq.~\eqref{eq:linear-degradation-model} and the corresponding energy minimisation based on the ``maximum a posteriori'' (MAP) method. In dictionary-based methods, denoising is performed under the assumption that the true signal $\bm{u}$ can be represented as a linear combination of the atoms in a dictionary matrix $\bm{D}$, which serve as the building blocks of the reconstructed signal. For a signal $\bm{u} \in \mathbb{R}^n$, the dictionary $\bm{D} = [\bm{d}_1, \ldots, \bm{d}_a] \in \mathbb{R}^{l \times a}$ is considered adapted to $\bm{u}$ if it can be approximated by a sparse combination of the columns of $\bm{D}$. In other words, there exists a sparse vector $\bm{\alpha} \in \mathbb{R}^a$ such that $\bm{u} \approx \bm{D}\bm{\alpha}$. This leads to the sparse decomposition problem,
\begin{equation}
    \bm{\alpha} = \arg\min_{\bm{\alpha}} \left\{ \frac{1}{2} \| \bm{h} - \bm{D}\bm{\alpha} \|_2^2 \right\} \quad \text{s.t.} \quad \psi(\bm{\alpha}),
\end{equation}
where $\psi$ is a prior that promotes sparsity in the solution. The convex Lagrangian relaxation of this ill-posed inverse problem is known as the LASSO (Least Absolute Shrinkage and Selection Operator)~\cite{Tibshirani:1996},
\begin{equation} \label{eq:SDL-lasso}
\bm{\alpha} = \arg\min_{\bm{\alpha}} \left\{ \frac{1}{2} \| \bm{D}\bm{\alpha} - \bm{h} \|_2^2 + \lambda \|\bm{\alpha}\|_1 \right\}.
\end{equation}

In this formulation, the first term is the \textit{fidelity term}, which quantifies how well the reconstruction matches the observed signal via the $\ell_2$-norm. The second term, referred to as the \textit{regularisation term}, encourages sparsity by promoting zeros in the vector coefficients $\bm{\alpha}$ and is weighted by the regularisation parameter $\lambda$. To avoid ambiguity in the context of denoising, we shall henceforth denote this parameter as $\lambda_{\text{den}}$, to distinguish it from other instances of $\lambda$ introduced later in this paper that may serve different purposes.

The value of $\lambda_{\text{den}}$ regulates the level of detail recovered in the sparse representation, and is therefore one of the main hyperparameters of our model.
The higher its value, the more weight is placed on the $\ell_1$-norm term, suppressing coefficients of $\bm\alpha$ and reducing the number of atoms used. Conversely, lower values prioritise the fidelity term, yielding reconstructions that closely match the data at the cost of using more (or even all) atoms. In the extreme case, the problem reduces to standard least squares.

To solve the LASSO problem we use the modified Least Angle Regression (LARS)~\cite{Efron:2004} algorithm. This algorithm provides a geometric-statistical procedure to obtain the entire solution path of the LASSO problem as the regularisation parameter $\lambda$ varies, enabling efficient exploration of sparsity levels. The algorithm is comparable to forward stepwise regression: it starts with all regression coefficients equal to zero and, at each iteration, identifies the predictor $u_i$ most correlated with the current residual. It then proceeds in that direction until another predictor reaches the same level of correlation. Rather than committing to one variable at a time, LARS continues in the equiangular direction between all equally correlated predictors, updating the coefficients jointly. The main advantage of this method is its efficiency when dealing with more predictors than observations ($a > l$), which is the case for overcomplete dictionaries. The interested reader is referred to~\cite{Efron:2004} for further details about the model.

\subsection{Dictionary learning problem\label{sec:dictionary-learning-problem}}

The performance of the denoising dictionary is enhanced by training it on the training dataset $\bm{U}$, from which $p$ patches\footnote{
    When reconstructing full signals, we refer to the overlapping patches into which they are divided as \emph{windows}. This distinction is purely semantic, and they will still be represented by the letter~$p$.
} of length $l$ are extracted, $\bm{U} = [\bm{u}_1, \dots, \bm{u}_p] \in \mathbb{R}^{l \times p}$, using a random-position windowing function. The training is performed by adding the dictionary matrix $\bm{D}$ as a variable in the minimisation problem,
\begin{equation} \label{eq:SDL-lasso-learning}
\bm{D} = \argmin_{\bm{\alpha},\bm{D}} \frac{1}{p} \sum_{i=1}^p \left\{
  \Vert \bm{u}_i - \bm{D}\bm{\alpha}_i \Vert_2^2 + \lambda\Vert\bm{\alpha}_i\Vert_1
\right\} ~,
\end{equation}
where $\bm{\alpha}_i$ denotes the $i$-th column of the coefficient matrix $\bm{\alpha} \in \mathbb{R}^{a \times p}$, which contains the sparse representations of the $p$ training patches. The dictionary atoms $(\bm{d}_j)_{j=1}^a$, with $\bm{d}_j \in \mathbb{R}^l$, are constrained to have $\ell_2$-norm less than or equal to one, $\bm{d}_j^T \bm{d}_j \leq 1$, to prevent $\bm{D}$ from becoming arbitrarily large.

The problem is solved using the algorithm proposed by Mairal et al. (2009)~\cite{Mairal:2009}, which includes mini-batch optimisation. It is a block-coordinate descent method that alternates between minimising $\bm{\alpha}$ and $\bm{D}$ at each iteration~$t$ (indicated with superscripts),
\begin{align}
\bm{\alpha}^t &= \argmin_{\bm{\alpha}} \left\{
  \frac{1}{2} \Vert \bm{U}^t - \bm{D}^{t-1} \bm{\alpha} \Vert_2^2
  + \lambda \Vert\bm{\alpha}\Vert_1
\right\}\,, \\
\bm{D}^t &= \argmin_{\bm{D}} \frac{1}{t} \sum_{i=1}^t \left\{
  \frac{1}{2} \Vert \bm{U}^i - \bm{D} \bm{\alpha}^i \Vert_2^2
  + \lambda \Vert\bm{\alpha}^i\Vert_1
\right\} ~.
\end{align}
This algorithm has the advantage of not requiring a learning rate. The regularisation parameter $\lambda$ used during training is conceptually different from the denoising parameter $\lambda_\text{den}$; we therefore refer to it as $\lambda_\text{learn}$ henceforth.

\subsection{Reconstruction quality assessment
\label{sec:reconstruction-quality-measures}}

To quantify the quality of the reconstruction we use the optimal regularisation parameter, $\lambda_{\text{opt}}$, defined as the one that gives the lowest value of a suitable error metric between the denoised and original signals. For completeness, we introduce three usual estimators, namely the Mean Squared Error (MSE), the Structural Similarity Index Measure (SSIM)~\cite{Wang:2004}, and the Match filter commonly used by the GW community~\cite{Turin:1960,Sathyaprakash:1991,Speri:2022}.

The MSE is typically defined as,
\begin{equation}
\text{MSE} (\bm{x}, \bm{y}) = \frac{1}{l} \sum_{i=1}^l (y_i - x_i)^2~,
\end{equation}
where $\bm{x}$ and $\bm{y}$ are the original and reconstructed signals respectively, and $l$ is the number of features. Throughout this work, we use the terms \textit{sample} and \textit{feature} in the machine-learning sense, referring respectively to a training instance and to its input variables. This avoids confusion with the signal-processing use of \textit{sample}, which denotes a single scalar value of a discretely sampled strain time series.

The SSIM, on the other hand, takes into account the structural information. It ranges between 1 (perfect correlation) and -1 (perfect anti-correlation), with 0 meaning no correlation at all, and is defined as,
\begin{equation}
\text{SSIM}(\bm{x}, \bm{y}) = \frac{(2\mu_x\mu_y + c_1)(2\sigma_{xy} + c_2)}{(\mu_x^2 + \mu_y^2 + c_1)(\sigma_x^2 + \sigma_y^2 + c_2)} ~,
\end{equation}
where $\mu_x$ ($\mu_y$) is the average of $\bm{x}$ ($\bm{y}$), $\sigma_x^2$ ($\sigma_y^2$) is the variance of $\mu_x$ ($\mu_y$), $\sigma_{xy}$ is the covariance between $\bm{x}$ and $\bm{y}$, and $c_1$ and $c_2$ are stability variables that depend on the dynamical range of the data.

The Match takes a step further than the SSIM; it provides a time-invariant frequency-wise similarity measure between the signals. It is defined as the correlation between the signals in frequency domain, maximised over a time shift~$\tau$\footnote{
    For complex signals, the complete expression also maximises over a global phase $\phi$: $(x | e^{i\phi} y_\tau)$. For real-valued signals, however, the operation is reduced to taking the absolute value, $|(x|y_\tau)|$.
},
\begin{equation}
    \mathcal{M} (\bm{x},\bm{y}) = \max_{\tau}\; \frac{
        \left| \langle \bm{x} \mid \bm{y}_\tau \rangle \right|
    }{
        \sqrt{ \langle\bm{x}\mid \bm{x}\rangle\, \langle\bm{y}\mid \bm{y}\rangle }
    }.
\end{equation}
Here, $\langle \;\cdot \mid \cdot\; \rangle$ denotes the PSD-weighted inner product which, for real-valued discrete signals, can be approximated by
\begin{equation}
    \langle \bm{x} \mid \bm{y} \rangle \approx 4\, \Re\!\left(
        \sum_k \frac{
            \tilde{x}_k\, \tilde{y}^{*}_k
        }{
            S_n(f_k)
        } \,\Delta f
    \right),
\end{equation}
where $\tilde{x}_k$ and $\tilde{y}_k$ are the discrete Fourier coefficients on positive frequencies $f_k$ with bin width $\Delta f$, and $S_n(f_k)$ is the detector's one-sided PSD.

\subsection{Classification\label{sec:classification}}

The problem of classification using SDL has been extensively addressed in the literature, with a large number of methods targeting different scenarios~\cite{Mairal:2008,Yang:2010,Zhang:2015,Xu:2017,Ngo:2022,Han:2024}. \textsc{clawdia} aims to gather effective classification methods specifically adapted to GW data analysis. However, ensuring code robustness requires exhaustive testing, which---due to the currently limited contributor base---requires incorporating one method at a time. As mentioned earlier, \textsc{clawdia} is a young project and currently implements only one classification method: the Low-Rank Shared Dictionary Learning (\textsc{LRSDL}) algorithm~\cite{vu-monga:2017}. This method was chosen first due to its relatively simple structure and high flexibility, achieved through the optional incorporation of shared feature components. The remainder of this section explains the core concept and mathematical foundation of \textsc{LRSDL}, which also serves as a general introduction to classification via SDL.

\begin{figure}[t]
    \centering
    \includegraphics[width=1\linewidth]{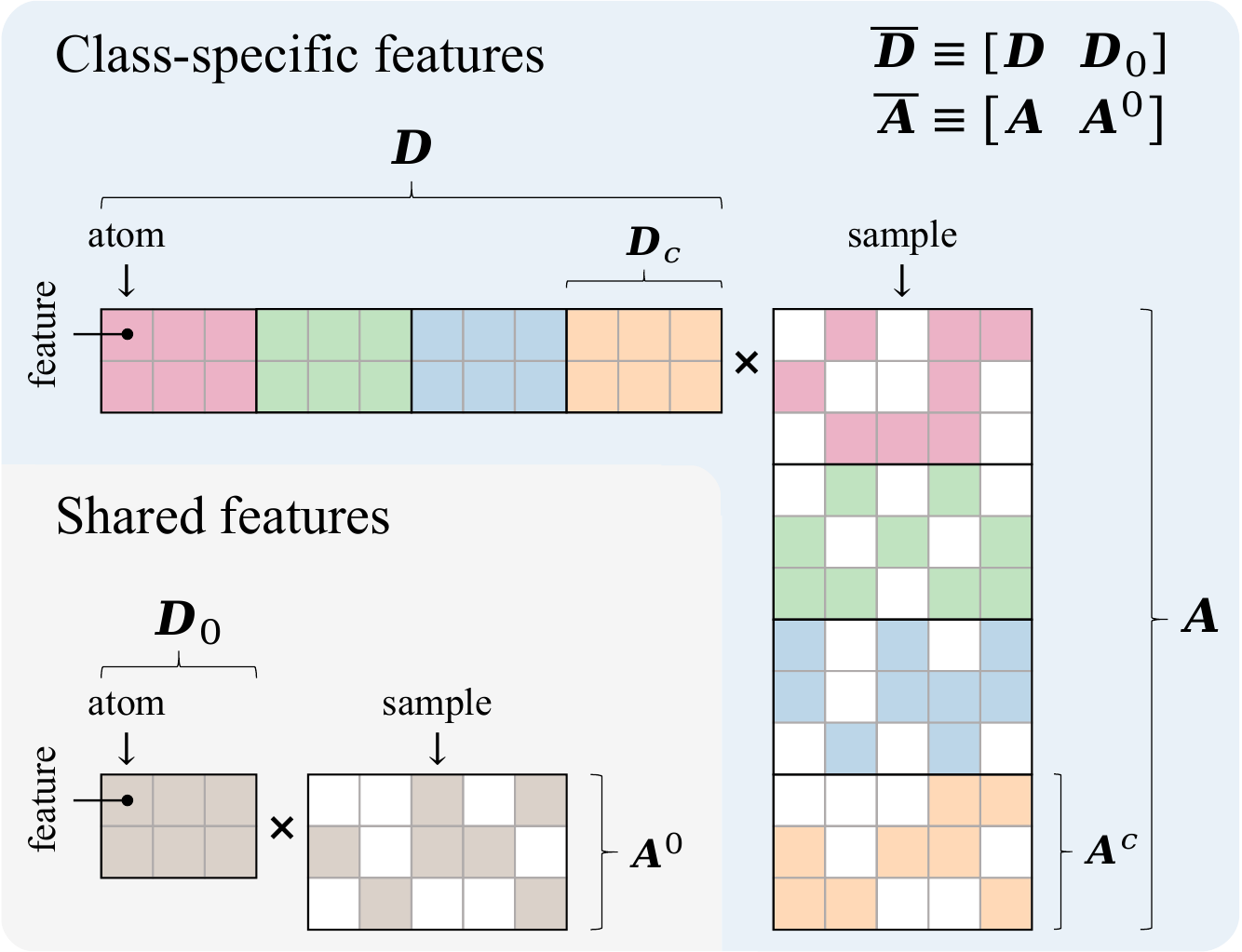}
    \caption{Schematic representation of the internal structure of the \textsc{LRSDL} classification dictionary $\bar{\bm{D}}$ and coefficient matrix $\bar{\bm{A}}$. In this example, five samples with two features each are reconstructed using the \textsc{LRSDL} model. The class-specific dictionary $\bm{D}$ and coefficient matrix $\bm{A}$ are formed by stacking four class-wise submatrices, each containing three atoms. The shared dictionary $\bm{D}_0$ and coefficient matrix $\bm{A}^0$ also contain three atoms, although their size do not need to match that of the class-specific components $\bm{D}_c$.}
    \label{fig:lrsdl-diagram}
\end{figure}

The \textsc{LRSDL} algorithm extends the Fisher Discrimination Dictionary Learning (FDDL)~\cite{yang:2011} method by capturing shared components among different waveform classes. It has demonstrated strong performance in scenarios with limited training data, outperforming previous dictionary-based algorithms in computational efficiency~\cite{vu-monga:2017}. The classification dictionary $\bar{\bm{D}}$ and the coefficient matrix $\bar{\bm{A}}$ have a more complex structure than those used for denoising, as illustrated in Figure~\ref{fig:lrsdl-diagram}. The dictionary consists of two semantically distinct parts: a set of class-specific dictionaries $\bm{D} \equiv \{\bm{D}_{c}\}_{c=1}^C$, each associated with a class, and a shared dictionary $\bm{D}_0$ that captures features common across multiple classes. We use the same notation for vector coefficients from the coefficient matrix $\bar{\bm{A}}$.

During training, the algorithm tries to reconstruct the input matrix $\bm{U}$ by linearly combining atoms from both types of dictionaries,
\begin{equation} \label{eq:lrsdl-reconstruction}
\bm{U} \approx \bm{D} \bm{A} + \bm{D}_0 \bm{A}^0 \,,
\end{equation}
which in Figure~\ref{fig:lrsdl-diagram} corresponds to the sum of the contributions from the blue (class-specific) and gray (shared) regions. The matrix $\bm{D}$ is obtained by horizontally stacking the class-specific dictionaries, $\bm{A} \equiv \{\bm{A}^c\}_{c=1}^C$ stacks the associated coefficient matrices, and $\bm{A}^0$ denotes the shared coefficient matrix. Different regularisation terms are imposed on each part of the model: the class-specific terms inherits FDDL-style constraints, while the shared dictionary $\bm{D}_0$ is enforced to be low-rank (via the nuclear norm) to prevent the assimilation of class-specific information. Furthermore, the shared coefficients in $\bm{A}^0$ are encouraged to be mutually similar, promoting a homogeneous contribution across classes. The complete cost function can be written as
\begin{equation}
\begin{aligned}
\bar{J}_{\bm{U}} (\bar{\bm{D}}, \bar{\bm{A}})
=&
\dfrac{1}{2} \bar{f}_{\bm{U}} (\bar{\bm{D}}, \bar{\bm{A}})
+
\lambda_1 \|\bar{\bm{A}}\|_1  \\
&+
\dfrac{\lambda_2}{2} \bar{g}(\bar{\bm{A}})
+
\eta \|\bm{D}_0\|_* \,,
\end{aligned}
\end{equation}
where $\bar{f}_{\bm{U}}$ is a discriminative fidelity term, $\bar{g}$ is the Fisher-based term extended to include shared components, and $\|\cdot\|_*$ denotes the nuclear norm~\cite{Recht:2010}. This leads to a total of six hyperparameters that govern the learning process. Three control the dictionary dimensions, $\bm{D}_c \in \mathbb{R}^{l \times a_C}$ and $\bm{D}_0 \in \mathbb{R}^{l \times a_0}$, while the other three are regularisation parameters: $\lambda_1$ (sparsity of $\bm{A}$), $\lambda_2$ (sparsity of $\bm{A}$ and homogeneity of $\bm{A}^0$), and $\eta$ (low-rank regularisation of $\bm{D}_0$).

The classification of an unknown input signal $\bm{u}$ is performed by first reconstructing it as in Eq.~\eqref{eq:lrsdl-reconstruction}, yielding a coefficient vector $\bar{\bm{\alpha}} = [\bm{\alpha}^T, (\bm{\alpha}^0)^T]^T$ with sparsity and homogeneity constraints on the shared part $\bm{\alpha}^0$. Then, the contribution of the shared dictionary is subtracted to isolate the class-specific content,
\begin{equation}
\bar{\bm{u}} = \bm{u} - \bm{D}_0 \bm{\alpha}^0 \,.
\end{equation}
Finally, the predicted class index is determined by identifying the class whose dictionary best matches $\bar{\bm{u}}$,
\begin{equation}
\argmin_{1 \leq c \leq C}
\left(
    w \| \bar{\bm{u}} - \bm{D}_c \bm{\alpha}^c \|_2^2
    + (1 - w) \| \bm{\alpha} - \bm{m}_c \|_2^2
\right) \,,
\end{equation}
where $\bm{m}_c$ is the mean coefficient vector of $\bm{A}^c$ obtained during training, and $w$ balances the contribution of reconstruction accuracy and proximity to the class mean in the coefficient space.

It should be noted that, due to the nature of the classification mechanism and the targeted signals, classification cannot currently be performed by segmenting the input into smaller windows. The classification atoms must span the entire duration of the meaningful content of the signal. This contrasts with denoising dictionaries, whose shorter atoms suffice to capture local features within the signal.

\section{FRAMEWORK AND FUNCTIONALITY\label{sec:CLAWDIA-framework-and-functionality}}

As outlined in the introduction, \textsc{clawdia} aims to integrate all data analysis stages that can be addressed by SDL-based algorithms in a unified system, with a particular emphasis on denoising, featuring various methods that not only support classification (and other future tasks) but can also function independently to produce high-quality signal reconstructions and noise removal.

In this section we provide a detailed description of \textsc{clawdia}, outlining its current individual components. Dedicated subsections address denoising algorithms, classification strategies, and parameter optimisation guidelines. Preprocessing steps and operations are performed using \textsc{gwadama} which, being extrinsic to the SDL methods, is briefly introduced in Appendix~\ref{sec:appendix-gwadama}.

\subsection{Denoising methods}

In \textsc{clawdia}, denoising methods are implemented through a structured hierarchy of Python classes, each corresponding to a specific dictionary model. These classes encapsulate the logic and parameters required to initialize, train, and apply a dictionary to GW data through a variety of dedicated methods. Each instance of such a class represents a single dictionary, ready to perform signal denoising.

This organisational strategy generalises beyond denoising: the same class-based framework is designed to support classification, and it will be naturally extended to other potential applications such as parameter estimation or detection. At present, the main criterion for grouping dictionaries is their back-end implementation, with future extensions allowing for a higher-level classification based both on their intended purpose and underlying implementation. For instance, dictionary classes designed for classification will be listed independently of those for denoising, even if both are based on the same SDL model. In such cases, inheritance will be used to define purpose-specific subclasses.

All current denoising methods are implemented within the \texttt{DictionarySpams} class, which leverages the LASSO solver provided by \textsc{SPAMS-python}~\cite{spams-python}. This choice is motivated by its robustness and extensive testing across diverse signal scenarios, while providing a reasonable computational performance. The implemented methods address specific denoising challenges, such as mitigating edge effects, tuning regularisation parameters, and allowing certain configurations to explicitly favour parallelisation over precision for large-scale datasets.

Importantly, denoising methods are designed with a dual purpose. On the one hand, they aim to produce signal reconstructions optimised for minimal residual noise, suitable for standalone use or direct analysis. On the other hand, they serve as an intermediate preprocessing step for subsequent analysis stages, such as classification, where preserving discriminative features or achieving computational efficiency may be prioritised over strict noise reduction.

This section provides an overview of the denoising strategies currently implemented in \textsc{clawdia}.

\subsubsection{Simple reconstruction
\label{sec:simple-reconstruction}}

The most straightforward denoising method implemented in \textsc{clawdia} is based on a sliding window approach, with its most simple form implemented as the \texttt{reconstruct} method. Given a generic signal $\bm{h} \in \mathbb{R}^L$, where $L > l$, the signal is divided into overlapping windows $\bm{p}^i$ of fixed length $l$ using a sliding window, defined as:
\begin{equation}
    \bm{p}^i = (h_{i\cdot s + k})_{k=1}^l
\end{equation}
where $s$ denotes the step size between consecutive windows, and $k$ is the index of the component of the $i$-th window. This ensures that every segment of the signal is covered by at least one window, with overlapping sections procuring smooth transitions in the final reconstruction.

Each window is reconstructed individually using the denoising dictionary $\bm{D}$, with a fixed regularisation parameter $\lambda_{\text{den}}$ controlling the sparsity of the reconstruction,
\begin{equation}
    \tilde{\bm{p}}^i = \bm{D} (\bm{p}^i; \lambda_{\text{den}})~.
\end{equation}
The atoms in $\bm{D}$ are trained to represent key features of GW signals while suppressing noise, ensuring robustness even under noise-dominated conditions.

After all windows are reconstructed, the denoised signal $\bm{r}$ is reassembled by averaging overlapping sections, minimising artifacts and preserving continuity. This can be expressed for each component $j$ of the reassembled reconstruction as,
\begin{equation}
    r_j = \dfrac{
        \displaystyle\sum\limits_{i = \lceil (j - l)/s \rceil}^{\lfloor (j-1)/s \rfloor} \tilde{p}^i_{j - i \cdot s + 1}
        }{
        \lfloor (j-1)/s \rfloor - \lceil (j - l)/s \rceil + 1
    }~,
\end{equation}
where $\lfloor \cdot \rfloor$ and $\lceil \cdot \rceil$ represent the floor and ceiling functions, respectively.
The summation includes only valid windows $i$ for which the denoised patch $p^i$ exists. That is, $i \in \{1, 2, \dots, N\}$, where $N$ is the number of windows extracted from the original signal.
The denominator accounts for the number of overlapping windows contributing to each component.

It must be noted that in order to avoid numerical errors, all split windows are normalised to their $\ell_2$ norm, matching that of the dictionary's atoms. To preserve their original relative signal energy, each denoised window is rescaled by the inverse of its respective original norm.

To optimise memory and CPU usage, \textsc{clawdia} includes a minibatch implementation, \texttt{reconstruct\_minibatch}, which takes full advantage of the parallelisation capabilities of the LASSO function in the \textsc{SPAMS-python} package. Instead of processing the overlapping windows from a single signal each time, windows from multiple signals are grouped into minibatches for reconstruction. The position and origin of each window is tracked, ensuring accurate reassembly into their respective signals. This approach allows handling large datasets, regardless of the number of signals or their length, while preserving continuity across overlapping windows.

This method is particularly useful for scenarios where simplicity and scalability are required. It is also used for other more advanced denoising methods as the core denoising step, supporting complex signal scenarios.

\subsubsection{Margin-constrained reconstruction
\label{sec:margin-constrained-reconstruction}}

The margin-constrained reconstruction method, named as \texttt{reconstruct\_margin\_constrained} in \textsc{clawdia}, is designed to perform denoising with minimal noise recovery as priority. Unlike standard approaches, which require manual tuning of the regularisation parameter $\lambda_\text{den}$, this method automatically determines it by minimising noise recovery at the specified strain interval (to a certain threshold) while ensuring non-zero denoising for the signal of interest, i.e., avoiding an over-regularisation that could distort or diminish it. This is achieved through the modified bisection search outlined in Algorithm~\ref{alg:CLAWDIA-margin-constrained-denoising}. The process begins by defining an objective function $f(\lambda)$, which measures the magnitude of the reconstruction at the margin,
\begin{equation}
    f(\lambda) = \left\| \text{Reconstruct}\left( (s_i)_{i=0}^{m-1}, \lambda \right) \right\|_1
\end{equation}
where $(s_i)_{i=0}^{m-1}$ represents the margin of size $m$ at the beginning of the signal array, and $\|\cdot\|_1$ denotes the $\ell_1$ norm, quantifying the total absolute reconstruction at the margin. The algorithm initializes the interval $[a, b]$ with the range of possible $\lambda$ values, ensuring that $f(a)$ and $f(b)$ are valid. The bisection search iteratively evaluates $f(\lambda)$ at the midpoint of the interval, updating the bounds until convergence is achieved based on the specified absolute and relative tolerances.

\begin{algorithm}[t]
    \setstretch{1.1}
    \KwIn{Signal $s \in \mathbb{R}^n$, margin size $m \in \mathbb{N}$, $\lambda$ range $[\lambda_{\min}, \lambda_{\max}]$, max iterations $N$, tolerances $\varepsilon_{\text{abs}}, \varepsilon_{\text{rel}}$}
    \KwOut{Final reconstructed signal}
    \BlankLine
    Define margin as $(s_i)_{i=0}^{m-1}$\;
    Define objective function $f(\lambda) = \| \texttt{Reconstruct}(s[:m], \lambda) \|_1$\;
    Initialize interval $[a, b] \gets [\lambda_{\min}, \lambda_{\max}]$\;
    \If{$f(a) = 0$}{
        Swap $a$ and $b$\;
    }
    \For{$i \gets 1$ \KwTo $N$}{
        Compute midpoint $x_m \gets (a + b)/2$\;
        Evaluate $f(x_m)$\;
        \eIf{$f(x_m) \neq 0$}{
            Update $a \gets x_m$\;
        }{
            Update $b \gets x_m$\;
        }
        \If{$|b - a| < \varepsilon_{\text{\rm abs}} + \varepsilon_{\text{\rm rel}} \cdot |x_m|$}{
            Set $\lambda_{\text{opt}} \gets x_m$\;
            Compute $\tilde{s} \gets \text{Reconstruct}(s, \lambda_{\text{opt}})$\;
            \Return $\tilde{s}$\;
        }
    }
    \caption{Margin-constrained reconstruction.}\label{alg:CLAWDIA-margin-constrained-denoising}
\end{algorithm}

This approach eliminates the need for manual tuning of $\lambda$, simplifying the reconstruction process. However, the computational cost of this method is higher than simpler techniques due to the multiple reconstructions required during the bisection search. Moreover, the bisection process makes the method not trivial to parallelise (although this is expected to be addressed in a future update). This trade-off between accuracy and computational expense must be carefully considered based on the specific application.

The margin-constrained reconstruction method is particularly advantageous in scenarios where the target signal power is sufficiently above the background noise level, providing a practical solution for denoising short GW signals.

\subsubsection{Iterative residual reconstruction}
\label{sec:iterative-reconstruction}

The iterative residual subtraction method is implemented in \textsc{clawdia}'s SPAMS-based dictionary as \texttt{reconstruct\_iterative}, inspired by the approach described in~\cite{Torres-Forne:2020}. This method reconstructs multiple signals by iteratively refining their residuals, reducing the reconstruction process's sensitivity to the regularisation parameter $\lambda$. Consequently, a single $\lambda$ value can be used across all signals in the dataset, simplifying parameter selection.

The procedure, summarized in Algorithm~\ref{alg:CLAWDIA-iterative-denoising}, begins by initializing the cumulative reconstruction $\bm{r}$ to zero and setting the residual $\bm{r}^{\text{res}}$ equal to the original signal $\bm{s}$. During the first iteration, the signal is reconstructed using the regularisation parameter $\lambda$, and the resulting reconstruction is subtracted from the original signal to compute the initial residual,
\begin{equation}
    \bm{r}^{\text{res}}_1 = \bm{s} - \text{Reconstruct}(\bm{s}, \lambda).
\end{equation}
Subsequent iterations refine the residual by reconstructing it and subtracting the output from itself. The cumulative reconstruction is updated at each step:
\begin{equation}
    \bm{r}^{\text{step}}_i = \text{Reconstruct}(\bm{r}^{\text{res}}_i, \lambda), \quad \bm{r} \gets \bm{r} + \bm{r}^{\text{step}}_i.
\end{equation}
The algorithm terminates when the Euclidean norm of the difference between consecutive residuals falls below a convergence threshold $\varepsilon$, or when the maximum number of iterations, $N$, is reached.

\begin{algorithm}[t]
    \setstretch{1.1}
    \KwIn{Signals $\bm{s}$, regularisation parameter $\lambda$, step size $s$, batch size $B$, maximum iterations $N$, convergence threshold $\varepsilon$}
    \KwOut{Final reconstructed signals $\bm{r}$}

    \BlankLine
    Initialize $\bm{r} \gets \bm{0}$, $\bm{r}^{\text{res}} \gets \bm{s}$, $\bm{r}^{\text{old}} \gets \bm{s}$\;
    Initialize iteration counter $n \gets 0$\;

    \While{convergence not met for all signals \textbf{and} $n < N$}{
        Compute reconstruction: $\bm{r}^{\text{step}} \gets \texttt{Reconstruct\_Batch}(\bm{r}^{\text{res}}, \lambda, B)$\;
        Update cumulative reconstruction: $\bm{r} \gets \bm{r} + \bm{r}^{\text{step}}$\;
        Update residual: $\bm{r}^{\text{res}} \gets \bm{r}^{\text{res}} - \bm{r}^{\text{step}}$\;
        Compute residual change: $\Delta \bm{r} \gets \|\bm{r}^{\text{res}} - \bm{r}^{\text{old}}\|_2$\;

        \If{$\Delta \bm{r} < \varepsilon$ for all signals}{
            Mark convergence as successful\;
        }
        \Else{
            Update $\bm{r}^{\text{old}} \gets \bm{r}^{\text{res}}$\;
            Increment $n \gets n + 1$\;
        }
    }
    \caption{Iterative residual reconstruction.}
    \label{alg:CLAWDIA-iterative-denoising}
\end{algorithm}

This approach avoids normalising the windows into which each signal is split. Normalisation would cause the residual amplitudes to be amplified iteratively, potentially slowing or even preventing convergence and distorting the reconstructed signal. This design choice ensures the robustness of the method against minor variations in signal characteristics, even when using a fixed $\lambda$ value.

There are several advantages to this method. By minimising the influence of individual signal characteristics, the algorithm facilitates the optimisation of a single $\lambda_\text{den}$, thereby simplifying parameter selection and ensuring consistent denoising performance across diverse datasets. Avoiding window normalisation further enhances the dictionary's discriminative ability, potentially yielding more accurate reconstructions. The method also scales to large-scale GW analysis via minibatch processing, which improves memory use and computational efficiency by processing groups of signals concurrently. At each iteration, reconstruction is performed on a minibatch, and the convergence criterion is evaluated collectively across all signals using vectorised operations. This approach simplifies the implementation and ensures that minibatches consistently contain the maximum number of signals, fully utilising the parallelisation capabilities of SPAMS's \texttt{lasso} function and substantially reducing computation time compared with per-signal processing.

Despite the inherent cost of such iterative method, this trade-off is often justified by its robust performance and its utility as a preprocessing step for subsequent tasks. Its demonstrated effectiveness in denoising transient GW signals for classification purposes, as illustrated in Section~\ref{sec:illustrative-examples-denoising} (see also~\cite{Torres-Forne:2020}), underscores its significance within the \textsc{clawdia} pipeline.

\subsubsection{Reference-guided reconstruction
\label{sec:reference-guided-reconstruction}}

The reference-guided reconstruction method, implemented in \textsc{clawdia} as \texttt{reconstruct\_loss\_optimised}, follows a similar approach to the margin-constrained method. Rather than requiring a predefined regularisation parameter, it dynamically selects the optimal value $\lambda_{\text{opt}}$ by minimising the reconstruction error with respect to a reference signal under a predefined error function (by default, the match measure defined in Section~\ref{sec:reconstruction-quality-measures}),
\begin{equation}
\lambda_{\text{opt}} = \arg \min_{\lambda} \left[ 1 - \text{Match}(\bm{h}_{\text{ref}}, \bm{h}_{\text{rec}}(\lambda)) \right].
\end{equation}

The optimisation is performed using SciPy's \texttt{minimise\_scalar} function over the logarithmic space of $\lambda$, improving the convergence speed. The procedure, detailed in Algorithm~\ref{alg:clawdia-optimum-reconstruct}, naturally yields high-fidelity reconstructions when reference signals accurately represent the desired outputs. While not primarily designed for practical denoising---mainly because of the frequently lack of reference signal---it is well-suited for hyper-parameter exploration. In particular, analyzing the distribution of $\lambda_{\text{opt}}$ across a population of known signals can offer valuable insights for parameter tuning and further studies in GW data analysis.

\begin{algorithm}[t]
    \setstretch{1.1}
    \KwIn{Signal $\bm{h}$, reference signal $\bm{h}_{\text{ref}}$, dictionary $\bm{D}$, bounds $[\lambda_{\min}, \lambda_{\max}]$}
    \KwOut{Denoised signal $\bm{h}_{\text{rec}}$ and optimal parameter $\lambda_{\text{opt}}$}

    \BlankLine
    Define reconstruction: $\bm{h}_{\text{rec}}(\lambda) \gets \texttt{Reconstruct}(\bm{h}, \bm{D}, \lambda)$\;
    Define loss: $\texttt{loss}(\lambda) \gets 1 - \text{SSIM}(\bm{h}_{\text{ref}}, \bm{h}_{\text{rec}}(\lambda))$\;

    \BlankLine
    Find $\lambda_{\text{opt}} \gets \argmin_{\lambda \in [\lambda_{\min}, \lambda_{\max}]} \texttt{loss}(\lambda)$\;
    Compute $\bm{h}_{\text{rec}} \gets \bm{h}_{\text{rec}}(\lambda_{\text{opt}})$\;

    \Return $\bm{h}_{\text{rec}}, \lambda_{\text{opt}}$\;

    \caption{Reference-guided reconstruction.}
    \label{alg:clawdia-optimum-reconstruct}
\end{algorithm}

\subsection{Classification with \textsc{LRSDL}\label{sec:classification-with-LRSDL}}

The classification method introduced in Section~\ref{sec:classification} is implemented in \textsc{clawdia} through the \texttt{DictionaryLRSDL} class, a wrapper around the original \textsc{LRSDL} implementation~\cite{DICTOL:2025} provided by the authors. This class adapts the method to the structure and conventions of the framework, ensuring compatibility with GW signals and providing a consistent interface for training, prediction, and saving.

Training is performed by the \texttt{fit} method, which accepts multi-class input arrays along with corresponding labels, and supports optional segmentation of each signal into overlapping windows. Low-energy windows can be excluded using a relative $\ell_2$-norm threshold, which helps filter out segments that may lack informative content when segmentation is enabled. Internally, the training data is sorted by class and checked to ensure that the number of windows per class suffices for the requested dictionary dimensions. The dictionary is then learned using the underlying \textsc{LRSDL} algorithm with class-specific and shared atoms, as described in Section~\ref{sec:classification}.

Once trained, the dictionary can be used for prediction through the \texttt{predict} method, which normalises the input signals, optionally crops them to match the atom length, and assigns a class label by identifying the class whose atoms best match the input according to the \textsc{LRSDL} loss function. A rejection threshold can be applied to the reconstruction loss, allowing uncertain predictions to be marked with a default unknown label.

The class includes a method for saving the trained model to disk in \textsc{NumPy}'s \texttt{.npz} format, and a module-level \texttt{load} function is provided to restore saved instances. A string-based representation is also available for inspecting the dictionary's parameters and internal state.

\subsection{Optimisation guidelines\label{sec:clawdia-optimisation-guidelines}}

The performance of \textsc{clawdia} depends critically on the configuration of its components, including the preprocessing pipeline, the structure and sparsity of its dictionaries, and the regularisation terms used during reconstruction or classification. While the framework currently supports denoising and classification, it is designed as a general-purpose tool for sparse modeling in GW data analysis. As such, this section presents general optimisation guidelines applicable to a range of applications, including future extensions such as signal detection or parameter estimation.

\subsubsection{General strategy}

Hyperparameters should be optimised with respect to the final task. For denoising, this typically involves minimising a reconstruction error metric such as MSE or SSIM. For classification, metrics such as the weighted F1-score are recommended to account for class imbalance and to balance precision and recall.

In general, all dictionaries and post-training parameters should be tuned toward the final objective as well. This unified approach has proven more effective than optimising individual modules in isolation, as shown in recent applications of \textsc{clawdia} to equation-of-state (EOS) inference from BNS postmergers~\cite{Llorens-Monteagudo:2025:bns}.

\subsubsection{Data splitting and cross-validation}

\textsc{clawdia} is designed with low-data regimes in mind, as these are common in current GW astrophysics. In such scenarios, a simple train/test split is often sufficient. For more robust performance evaluation, cross-validation is strongly recommended. K-fold cross-validation with stratification (in cases of class imbalance) has shown good performance in several related classification studies~\cite{Powell:2024,Llorens-Monteagudo:2025:bns}.

\subsubsection{Preprocessing configuration}

Preprocessing steps such as normalisation, resampling, whitening, and filtering are powerful tools that often enhance dictionary performance. These operations should be applied consistently across clean and noise-injected datasets to ensure that the training data reflect feature distributions comparable to the test data. This alignment improves the generalisation of learned representations to realistic scenarios.

Whitening is particularly important, as it suppresses spectral biases induced by the detectors' sensitivity curves and by long-duration transient noise. However, it must be applied with care: the whitening operation effectively acts as a frequency-dependent weighting, amplifying or suppressing different regions of the spectrum. Improper whitening can introduce bias by artificially boosting high-noise regions, skewing the learned dictionary toward uninformative features.

Highpass filtering is another important preprocessing tool, often beneficial across all applications to suppress low-frequency trends that could otherwise be misinterpreted as signal content. This is particularly relevant in time-domain applications of SDL, where louder spectral components tend to dominate the learned representations. More broadly, preprocessing decisions should reflect the target signals' frequency range, duration, and noise environment.

These operations have been implemented in \textsc{gwadama}, facilitating the homogeneous treatment between the training and test signals (see Appendix~\ref{sec:appendix-gwadama}).

\subsubsection{Dictionary training}

Training on noise-free or high-SNR signals (when available) has consistently led to better performance in both denoising and classification tasks. However, this recommendation comes with caveats. The current evidence is limited to time-domain signals, and it remains to be tested whether the same holds for frequency-domain or spectrogram-based representations. In addition, alternative objectives such as detection or parameter estimation may require different training strategies.

Most dictionaries in \textsc{clawdia} are trained using LASSO-based sparse coding. Across most SDL models, key hyperparameters include the atom length $l$, the number of atoms $a$, and the regularisation parameter $\lambda_{\text{learn}}$ that controls sparsity during training. Among these, the atom length $l$ is the most tightly bound to the physical properties of the target signals. For standard models such as LASSO, which lack explicit spectral constraints, the atom length effectively acts as a bandpass filter: longer atoms emphasize low-frequency structures, while shorter atoms prioritize sharper or higher-frequency features. It is therefore advisable to begin hyperparameter optimisation with $l$, as it plays a dominant role in determining the dictionary's capacity to capture relevant morphological features.

In contrast, the behavior of $a$ is usually straightforward: performance generally improves with increasing $a$ until it saturates. Overcomplete dictionaries, where the number of atoms exceeds their length ($a > l$), are generally recommended, as they provide more flexibility and redundancy in representing signal variability. However, this comes at the cost of increased computational demands and may require larger training datasets, particularly for larger $l$. For this reason, $a$ should be selected to balance reconstruction quality with computational efficiency, constrained by the available data and problem complexity.

\subsubsection{Classification optimisation}

The current classification model in \textsc{clawdia} is \textsc{LRSDL}. This algorithm is particularly effective when the target classes exhibit shared morphological components. It distinguishes between class-specific and shared dictionaries and imposes different constraints on each.

The classifier introduces several hyperparameters: the number of atoms per class ($a_C$), the number of shared atoms ($a_0$), and three regularisation terms: $\lambda_1$, $\lambda_2$, and $\eta$.
Before optimisation, users should assess whether the target classes exhibit meaningful shared structure. If not, disabling the shared dictionary ($a_0 = 0$) is often beneficial, as its inclusion may otherwise absorb class-discriminative features and degrade performance. If shared features are expected, we recommend the following progressive fixing strategy:
\begin{enumerate}
    \item Disable the shared dictionary.
    \item Tune $a_C$ and $\lambda_1$.
    \item Enable the shared dictionary.
    \item Optimise $a_0$, $\lambda_2$, and $\eta$.
\end{enumerate}
This strategy simplifies the high-dimensional search space and allows for more interpretable tuning decisions. It has shown robust performance in low-data regimes with limited computational resources.

Note that the \textsc{LRSDL} implementation in \textsc{clawdia} currently requires fixed-length input signals. The atom length $l$ must therefore be large enough to accommodate the full signal duration.

\subsubsection{Final considerations}

To summarize, we offer the following general recommendations:
\begin{itemize}
    \item Optimise all components of the pipeline with respect to the final objective (e.g., denoising, classification, detection).
    \item Normalise and whiten inputs to suppress detector-induced biases.
    \item Apply consistent preprocessing across clean and noise-injected signals.
    \item Use stratified K-fold cross-validation to make the most of limited training data.
    \item Use overcomplete dictionaries when data and resources allow.
    \item Train dictionaries on noise-free data unless task-specific evidence suggests otherwise.
    \item For classification, tune all parameters in low-SNR conditions to ensure robustness.
\end{itemize}

Further examples and suggested optimisation workflows will be available in the public repository and documentation of \textsc{clawdia}. Users are encouraged to adapt these recommendations to the characteristics of their data and application domain.

\section{CLASSIFICATION PIPELINE\label{sec:classification-pipeline}}

\begin{figure}
    \centering
    \includegraphics[width=\linewidth]{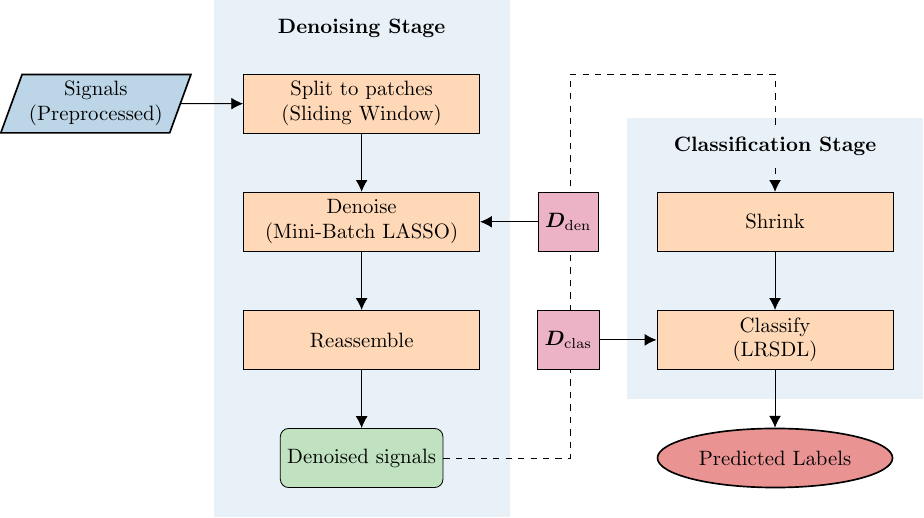}
    \caption{Workflow of the classification pipeline implemented in \textsc{clawdia}. Input signals are segmented into overlapping patches and denoised using the reconstruction dictionary $\bm{D}_\text{den}$ before being reassembled. The denoised signals are then classified using the dictionary $\bm{D}_\text{clas}$ trained for label discrimination.
    \label{fig:clawdia-flowchart}}
\end{figure}

The classification pipeline in \textsc{clawdia} is implemented as a lightweight Python class that assembles essential components of the framework into a workable sequence for supervised classification. It is currently intended as a reference implementation and a template for constructing SDL-based workflows, rather than the architectural centre of \textsc{clawdia}.

The pipeline integrates two main stages: preprocessing (denoising) and classification. Input signals may be conditioned by normalisation, whitening, resampling, and filtering; cropping can be applied to reduce computational cost prior to whitening. Denoising is performed via sparse reconstruction over overlapping patches using a pre-trained dictionary. The resulting signals are then passed to the classification stage, which assigns a class label based on the sparse decomposition with respect to a classification dictionary. A summary is shown in Figure~\ref{fig:clawdia-flowchart}.

Components are parameterised and can be replaced with alternative implementations. The denoising step supports multiple reconstruction strategies, including margin-constrained and iterative residual methods. The classification stage currently relies on the \textsc{LRSDL} model (discussed in Section~\ref{sec:classification-with-LRSDL}). The interface is kept simple to accommodate additional models with minimal changes. Preprocessing routines are provided via the complementary \textsc{gwadama} package.

The dictionaries are currently assumed to be trained and validated externally. The present release does not include built-in tools for dictionary optimisation; future updates may introduce routines to aid this, following the same modular and standardised design principles.

Although not the conceptual centre of \textsc{clawdia}, the pipeline provides a reproducible and extensible scaffold for SDL-based classification under realistic GW noise conditions. Potential extensions include detection and regression pipelines following the same design pattern.

\section{ILLUSTRATIVE RESULTS\label{sec:illustrative-examples}}

This section presents a set of illustrative examples of \textsc{clawdia}'s main functions. The aim is not to provide a quantitative performance evaluation, but rather to demonstrate how the framework integrates SDL methods into GW data analysis. For this reason, the datasets, hyperparameters, and training configurations are chosen ad-hoc focusing on the visual clarity of the results. An actual astrophysical application of \textsc{clawdia} to the classification of the equation of state of neutron stars through the reconstruction of simulated post-merger GW signals from BNS mergers is reported in~\cite{Llorens-Monteagudo:2025:bns}.

\subsection{Denoising of signal GW170817
\label{sec:illustrative-examples-denoising}}

\begin{figure}
    \centering
    \includegraphics[width=\linewidth]{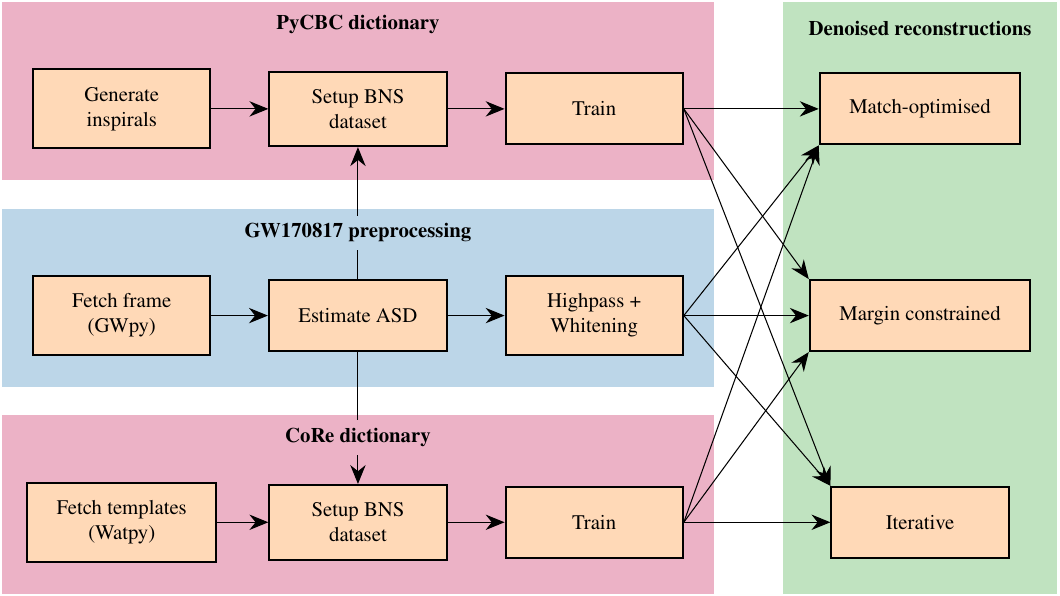}
    \caption{Workflow of the denoising of GW170817 by three different methods (green area). The blue area corresponds to the preprocessing steps applied to the input strain, including the estimation of the ASD from the detector's background noise, used to whiten the training signals by which the denoising dictionaries (pink areas) were trained.}
    \label{fig:workflow-denoising-gw170817}
\end{figure}

To illustrate the capabilities of the denoising methods, we use the \texttt{DictionarySPAMS} model to reconstruct the BNS merger signal GW170817 as observed by the LIGO Hanford detector. The data are obtained from the Gravitational-Wave Open Science Center (GWOSC)~\cite{GWOSC_O1_O2_2021,GWOSC_O3_2023}. To facilitate the visualisation of the steps followed, Figure~\ref{fig:workflow-denoising-gw170817} displays a diagram of the denoising workflow.

The training datasets are built using \textsc{gwadama}'s dataset class \texttt{UnlabeledWaves}, which exposes the preprocessing utilities with a simplified environment for unlabeled data.
For performance reasons, all data are downloaded, generated, or resampled, depending on their origin, with a constant sampling rate of 4096~Hz. This leaves an ample margin between the highest frequential features observed by the Hanford detector and the Nyquist frequency.
Furthermore, to reduce low-frequency noise all data are highpassed at 30~Hz.

First (blue region in Fig.~\ref{fig:workflow-denoising-gw170817}) we download 200 seconds of data with the GW170817 event at the centre using \textsc{GWpy}. The 99-second segment between one second after the event and the end of the frame is used to estimate the amplitude spectral density (ASD) of the background noise. With it, the GW170817 frame is whitened in order to highlight the GW features and get rid of (most of) the spectral lines. Having the strain of the event ready to be denoised, we set up two denoising dictionaries (pink regions in Fig.~\ref{fig:workflow-denoising-gw170817}) using \textsc{clawdia}'s \texttt{DictionarySPAMS} class. For both of them, all training waveforms are projected on the sky at the estimated position of the optical counterpart SSS17a, $
    \mathrm{RA} = 13^{\mathrm{h}}\,09^{\mathrm{m}}\,48.085^{\mathrm{s}},
    ~
    \mathrm{Dec} = -23^{\circ}\,22^{\prime}\,53.343^{\prime\prime}\ \text{(J2000)}
$~\cite{Coulter:2017}. Additionally, to simulate how these GWs would be observed by the Hanford detector, we whiten them using the estimated ASD of the detector's background noise, applying the same 30~Hz highpass filter.

The first dictionary (top pink region) is trained on \textsc{PyCBC}-generated GWs using the \texttt{PhenomPNRT} model\footnote{
    The complete name of \texttt{PhenomPNRT} in \textsc{LALSuite} (and by extension in \textsc{PyCBC}) is \texttt{IMRPhenomPv2\_NRTidal}~\cite{lalsimulation:IMRPhenomPv2_NRTidal}.
}~\cite{Dietrich:2019:PhenomPNRT} over a small range of total masses ($m \in [2.68, 2.78]\,\text{M}_\odot$) in order to provide variability in the training data, while the rest of parameters are taken directly from~\cite{LIGOVirgo:2019}. This dictionary is intended to accurately recover the inspiral region of GW170817, while deviations in the merger region are to be expected.
The overcomplete dictionary is composed of 800 atoms with 256 features each, and trained with $\lambda_\text{learn} = 0.2$ for 1000 iterations, values chosen intuitively from knowledge of previous works.

For completeness, we try to reconstruct the merger region of the GW with a second dictionary (bottom pink region) trained with CoRe waveforms from NR simulations including the same five different EOS used in~\cite{Llorens-Monteagudo:2025:bns}. In our example, however, the waveforms are trimmed, leaving all the available inspiral and up to a few milliseconds after merger. Moreover, instead of filtering out the prompt-collapse models, we remove the long-lived remnants, as there was no evidence of such component in the detected GW signal~\cite{GW170817}. This second dictionary is built with 750 atoms of 512 features each, twice the length of the \textsc{PyCBC} dictionary and enough to fit most NR waveforms whole. The reason behind this choice is to indirectly guide the dictionary to account for the global morphology of the CoRe waveforms, which we consider necessary given their stochastic and complex nature around the merger part, as opposed to the \texttt{PhenomPNRT} waveforms.
The dictionary is trained with the same $\lambda_\text{learn}$ and number of iterations as the \textsc{PyCBC} dictionary.

\begin{figure*}
    \centering
    \begin{subfigure}[b]{0.48\textwidth}
        \caption{Hanford strain}
        \includegraphics[width=\linewidth]{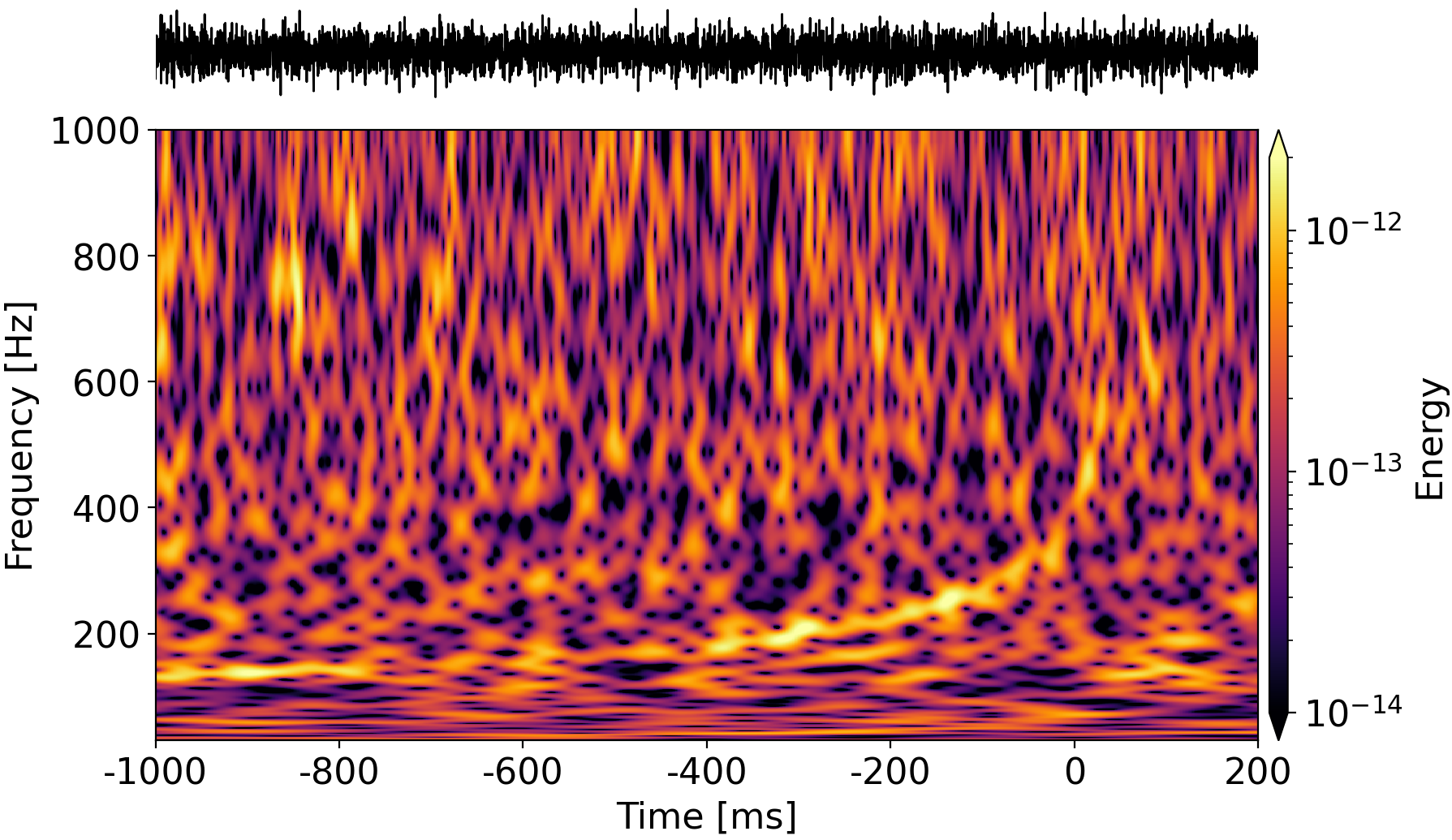}
        \label{fig:gw170817_denoisings:input}
    \end{subfigure}

    \vspace{-1em}

    \begin{subfigure}[b]{0.48\textwidth}
        \caption{Match-optimised reconstruction}
        \includegraphics[width=\linewidth]{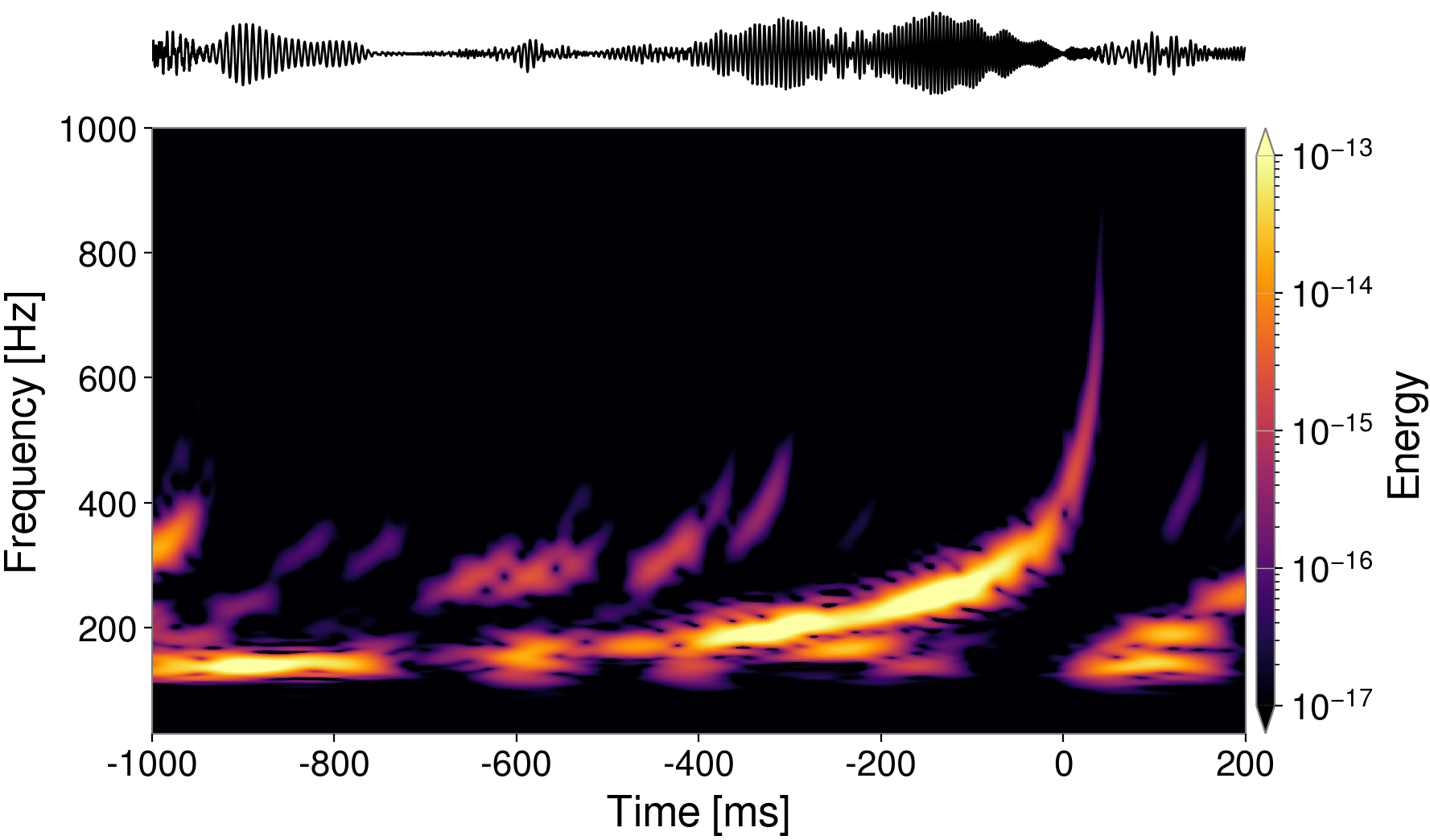}
        \label{fig:gw170817_denoisings:match_optimised_rec}
    \end{subfigure}\hfill
    \begin{subfigure}[b]{0.48\textwidth}
        \caption{Residual}
        \includegraphics[width=\linewidth]{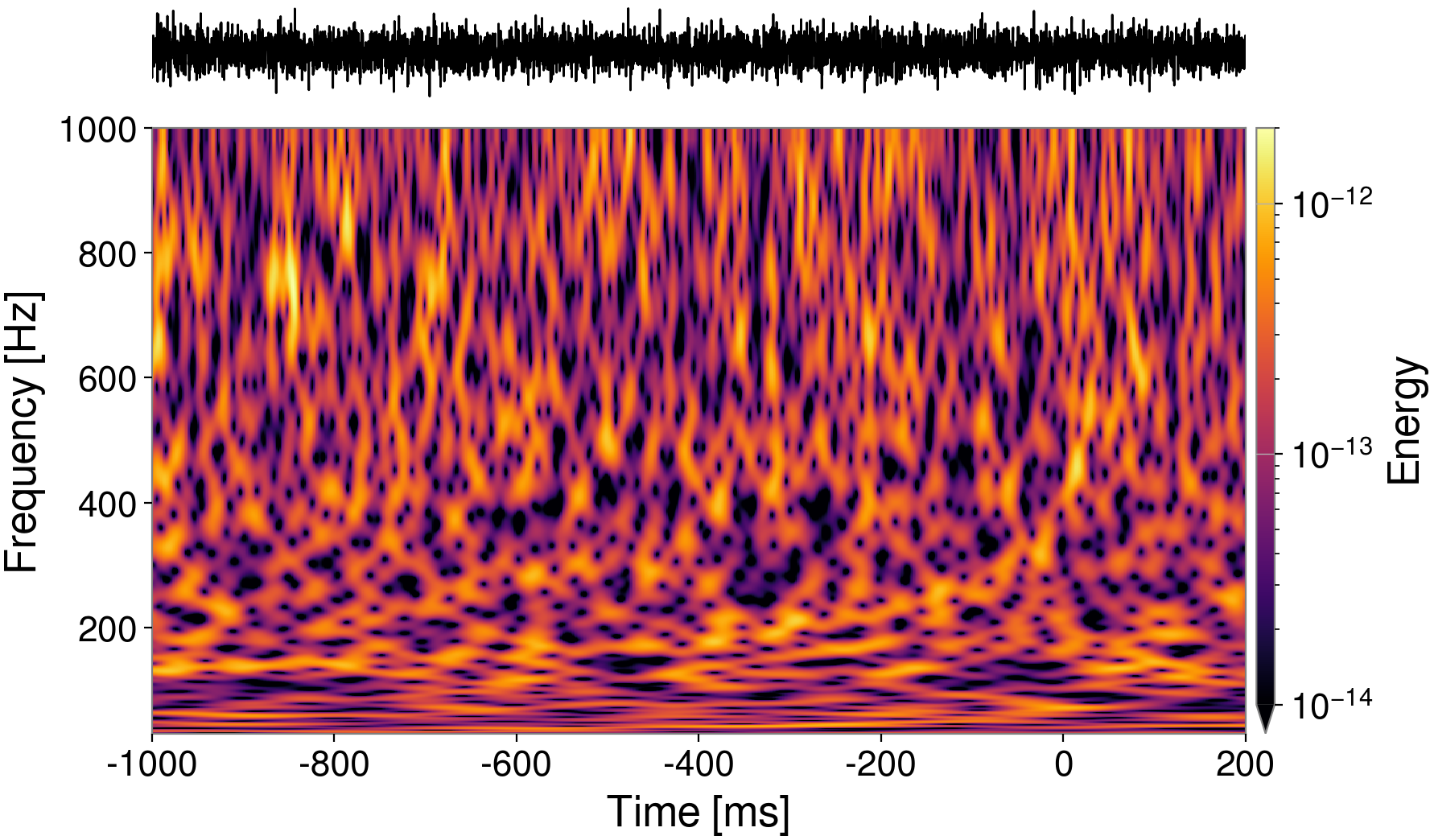}
        \label{fig:gw170817_denoisings:match_optimised_res}
    \end{subfigure}

    \vspace{-1em}

    \begin{subfigure}[b]{0.48\textwidth}
        \caption{Margin-constrained reconstruction}
        \includegraphics[width=\linewidth]{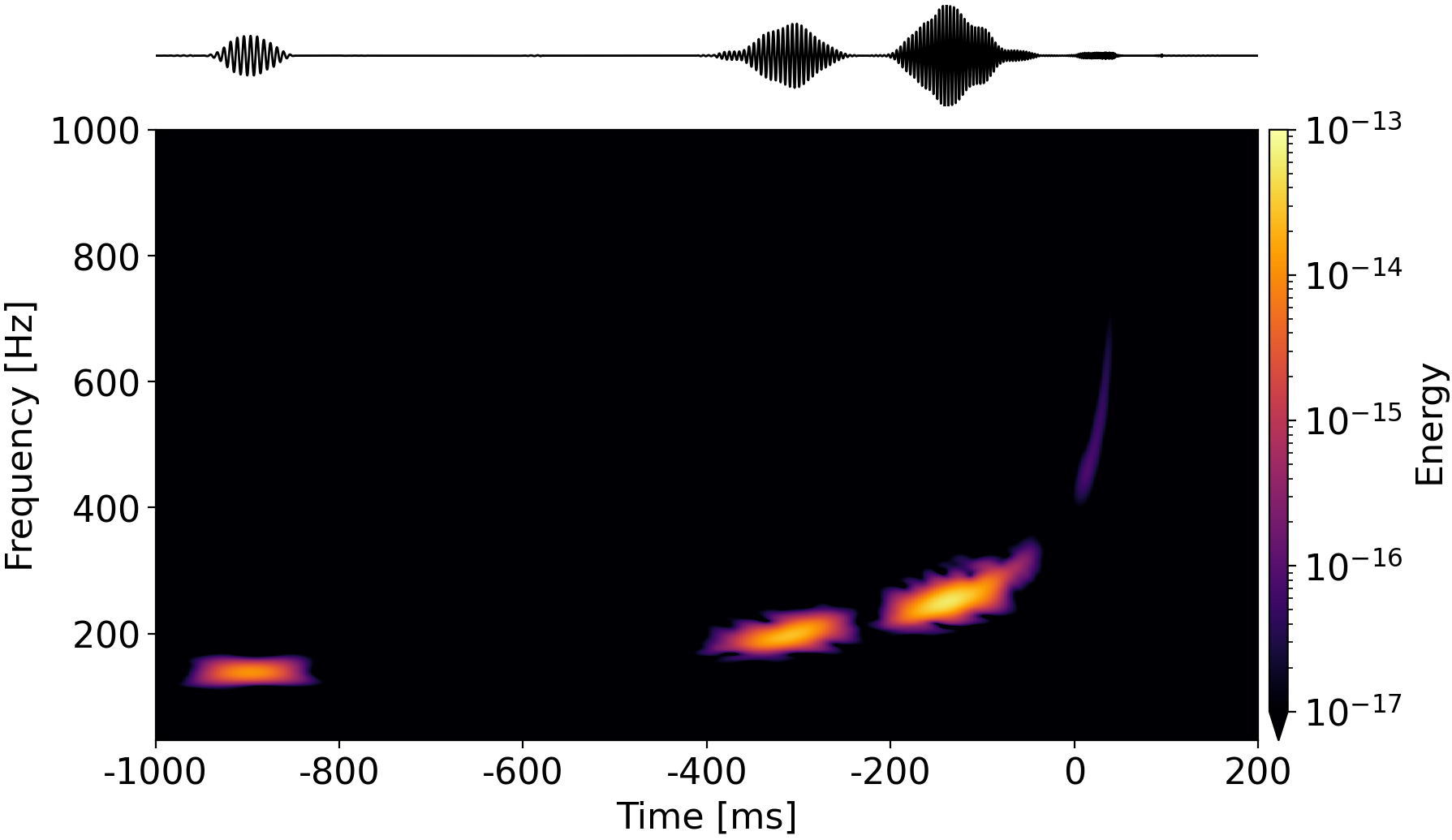}
        \label{fig:gw170817_denoisings:margin_constrained_rec}
    \end{subfigure}\hfill
    \begin{subfigure}[b]{0.48\textwidth}
        \caption{Residual}
        \includegraphics[width=\linewidth]{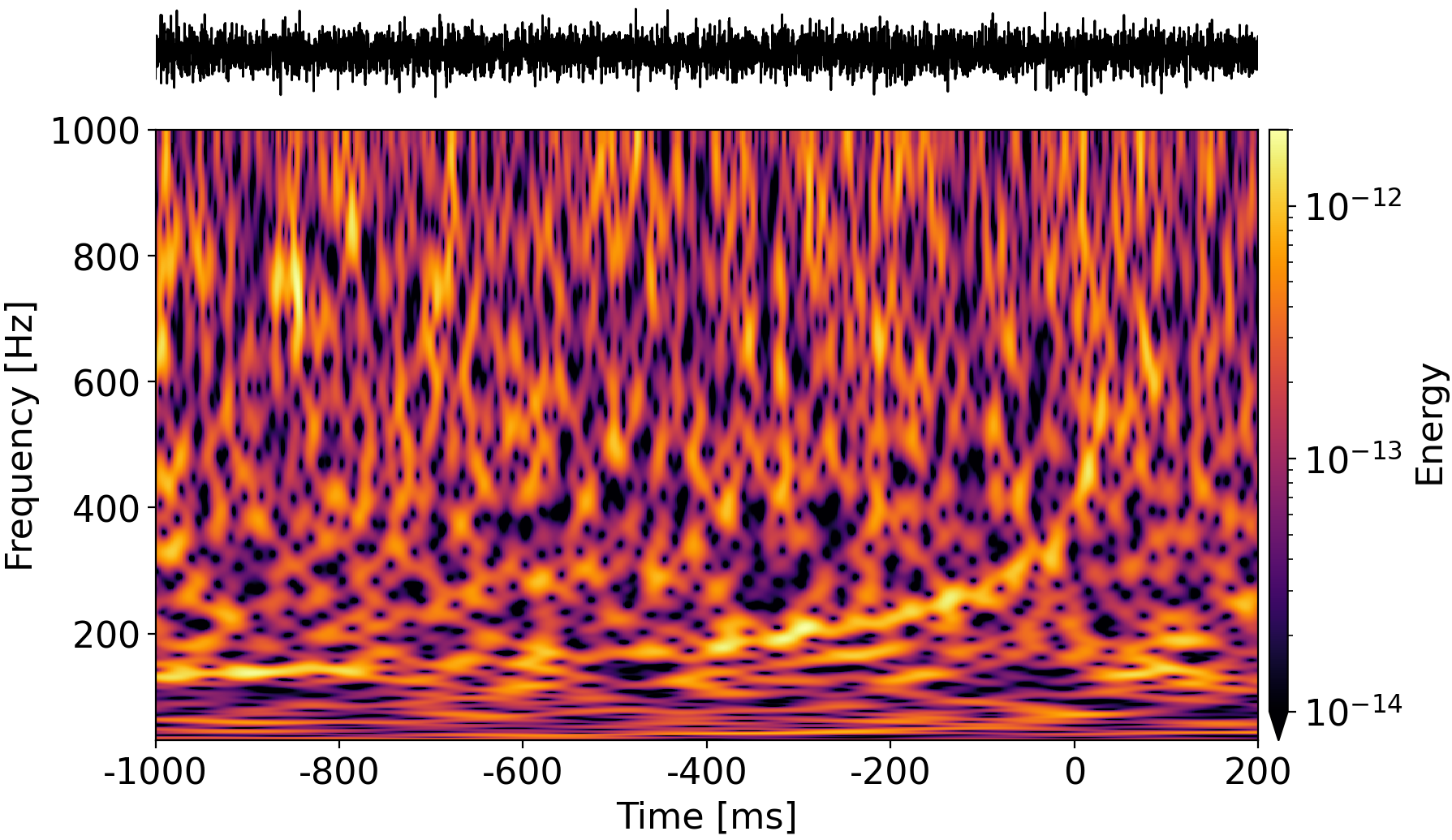}
        \label{fig:gw170817_denoisings:margin_constrained_res}
    \end{subfigure}

    \vspace{-1em}

    \begin{subfigure}[b]{0.48\textwidth}
        \caption{Iterative reconstruction}
        \includegraphics[width=\linewidth]{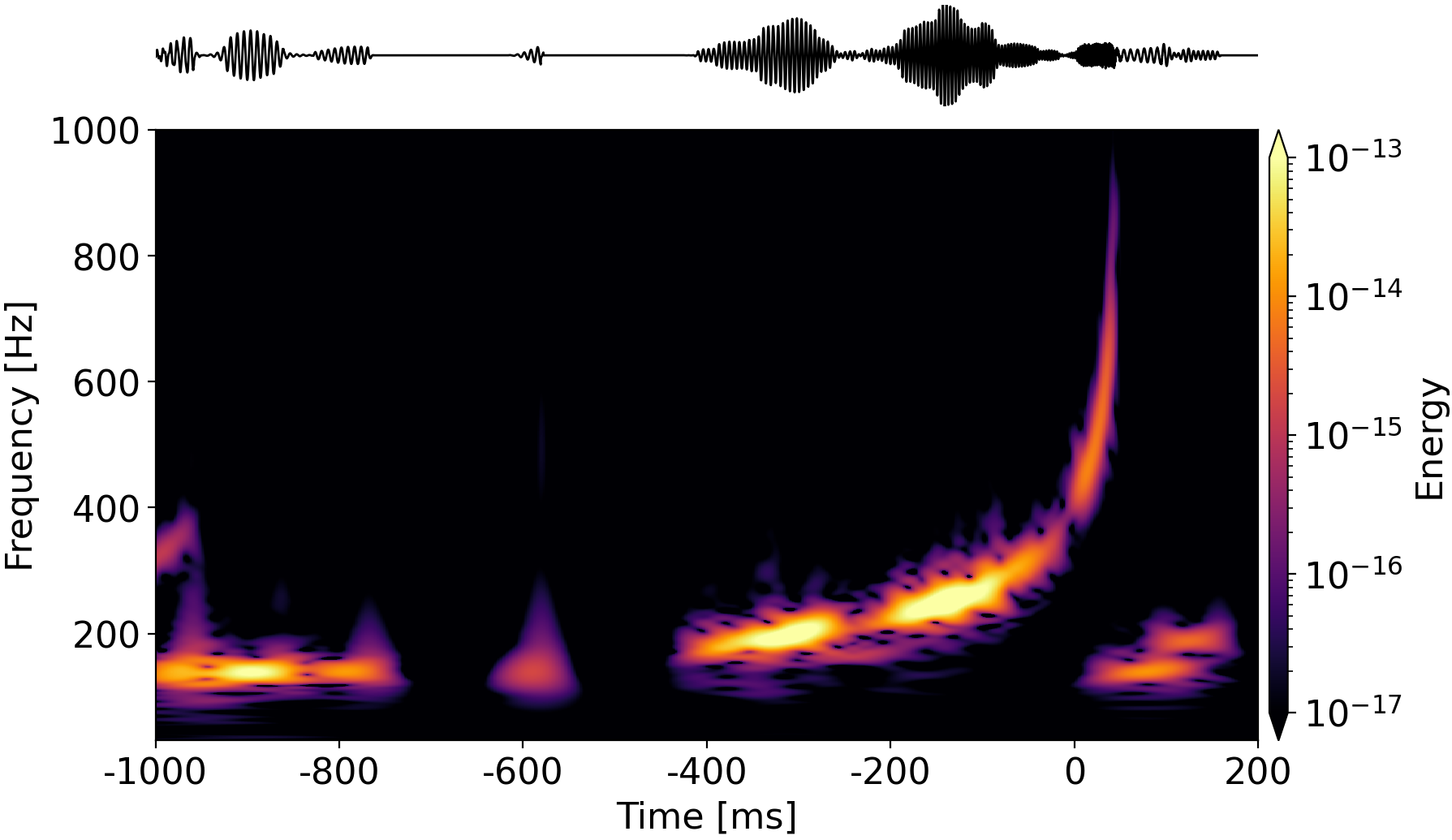}
        \label{fig:gw170817_denoisings:iterative_rec}
    \end{subfigure}\hfill
    \begin{subfigure}[b]{0.48\textwidth}
        \caption{Residual}
        \includegraphics[width=\linewidth]{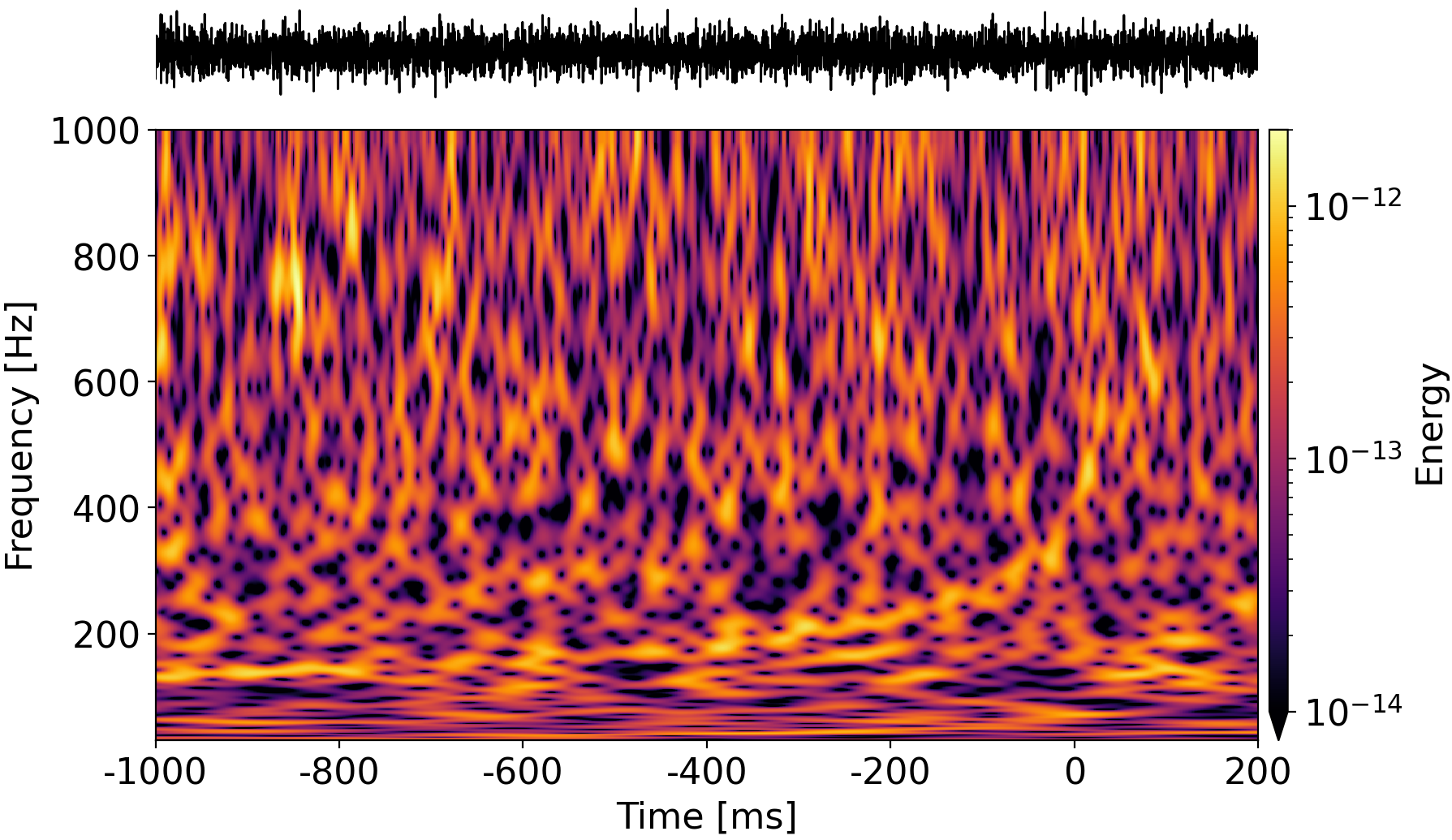}
        \label{fig:gw170817_denoisings:iterative_res}
    \end{subfigure}

    \caption{Time-frequency representations of the Hanford strain (\ref{sub@fig:gw170817_denoisings:input}) and, for each method, the reconstruction (left column) with its residual (right column). Residuals share the reference colour scale; reconstructions share a wider one for readability. A normalised time-series overlay (black) is shown above each panel.}
    \label{fig:gw170817_denoisings}
\end{figure*}

The denoising of GW170817 is completed by a naive linear combination of the reconstructions yielded by the two dictionaries, both applied with the same reconstruction method.
We repeat this process for three denoising methods: reference-guided, margin-constrained, and iterative residual.
The results are displayed in the time-frequency panels of Figure~\ref{fig:gw170817_denoisings}, with the normalised time-series representation of the strain on top of each panel.
The original Hanford strain is shown in Figure~\ref{fig:gw170817_denoisings:input} for reference, with the $\sim 8$~SNR GW slightly visible as a horizontal spectral feature that begins at around 150~Hz in the chosen time window, and sweeps up to 800~Hz a few milliseconds after the merger, which is set as the time origin.
The denoisings are organised on the left column, and the residuals on the right column. Neither the denoisings nor the reconstructions have been normalised after being computed in order to allow direct comparison between methods. Instead, the limits of the logarithmic colour scale that indicate the spectral energy density have been adjusted in two ranges: a two-magnitude high-energy range for the original strain and residuals, and a four-magnitude lower-energy range for the denoised strains. This choice increases the contrast in the residual spectra, since by construction they hold much more data than the denoised counterparts, while allowing the visualisation of fainter morphological details in the denoised strains. It must be noted, however, that the latter would not be needed in practical scenarios. Depending on the method or methods employed, denoised strains may be (and often are) scaled up to compensate for the global amplitude damping that these methods often display.

With the first denoising method, namely the match-optimised reconstruction (introduced in Section~\ref{sec:reference-guided-reconstruction}), we attempt to find the value of $\lambda_\text{den}$ that maximises the match between the dictionary reconstruction of the Hanford strain and the \textsc{PyCBC} waveform obtained with the estimated parameters for GW170817. In other words, we use the latter as a reference guide for both dictionaries. The result is a relatively aggressive reconstruction ($\lambda_\text{den}^\text{PyCBC} \approx 0.16$, $\lambda_\text{den}^\text{CoRe} \approx 0.19$) that recovers most of the astrophysical GW components (Figure~\ref{fig:gw170817_denoisings:match_optimised_res}) at the expense of leaking part of the background noise of the surrounding frequency bands alongside some artefacts (Figure~\ref{fig:gw170817_denoisings:match_optimised_rec}). Nevertheless, most of the areas in the time-frequency representation that fall relatively far from the GW signal have been effectively cleaned.

The second denoising method is the margin-constrained reconstruction, in which limiting the background noise is prioritised over signal recovery. The segment selected for minimisation begins 0.2 seconds after the merger, and extends for 0.5 seconds of what we would expect to be strictly detector noise. The result is a much more sparse reconstruction ($\lambda_\text{den}^\text{PyCBC} \approx 0.25$, $\lambda_\text{den}^\text{CoRe} \approx 0.21$) that discarded most of the background noise (Figure~\ref{fig:gw170817_denoisings:margin_constrained_rec}) at the expense of a partial recovery (Figure~\ref{fig:gw170817_denoisings:margin_constrained_res}) that misses some segments of the GW. It is worth noting, however, that in this much more discriminant configuration neither the high-frequency artefacts from the inspiral nor the horizontal lines after the merger appear.

The third denoising method is the iterative residual reconstruction. As explained in Section~\ref{sec:iterative-reconstruction}, it provides much more discriminant reconstructions with reduced loss in signal recovery thanks to its iterative denoisings subtracting relevant data from the residual. In fact, for this application we set both dictionaries to be much more discriminant than in the margin-constrained denoising ($\lambda_\text{den}^\text{PyCBC} \approx 0.9$, $\lambda_\text{den}^\text{CoRe} \approx 1.1$), yet most of the GW features captured by the match-optimised method are successfully recovered (Figure~\ref{fig:gw170817_denoisings:iterative_res}) while filtering most of the noise artefacts (Figure~\ref{fig:gw170817_denoisings:iterative_rec}).

\begin{figure*}
    \centering
    \begin{subfigure}[b]{0.48\textwidth}
        \caption{Blip}
        \includegraphics[width=\linewidth]{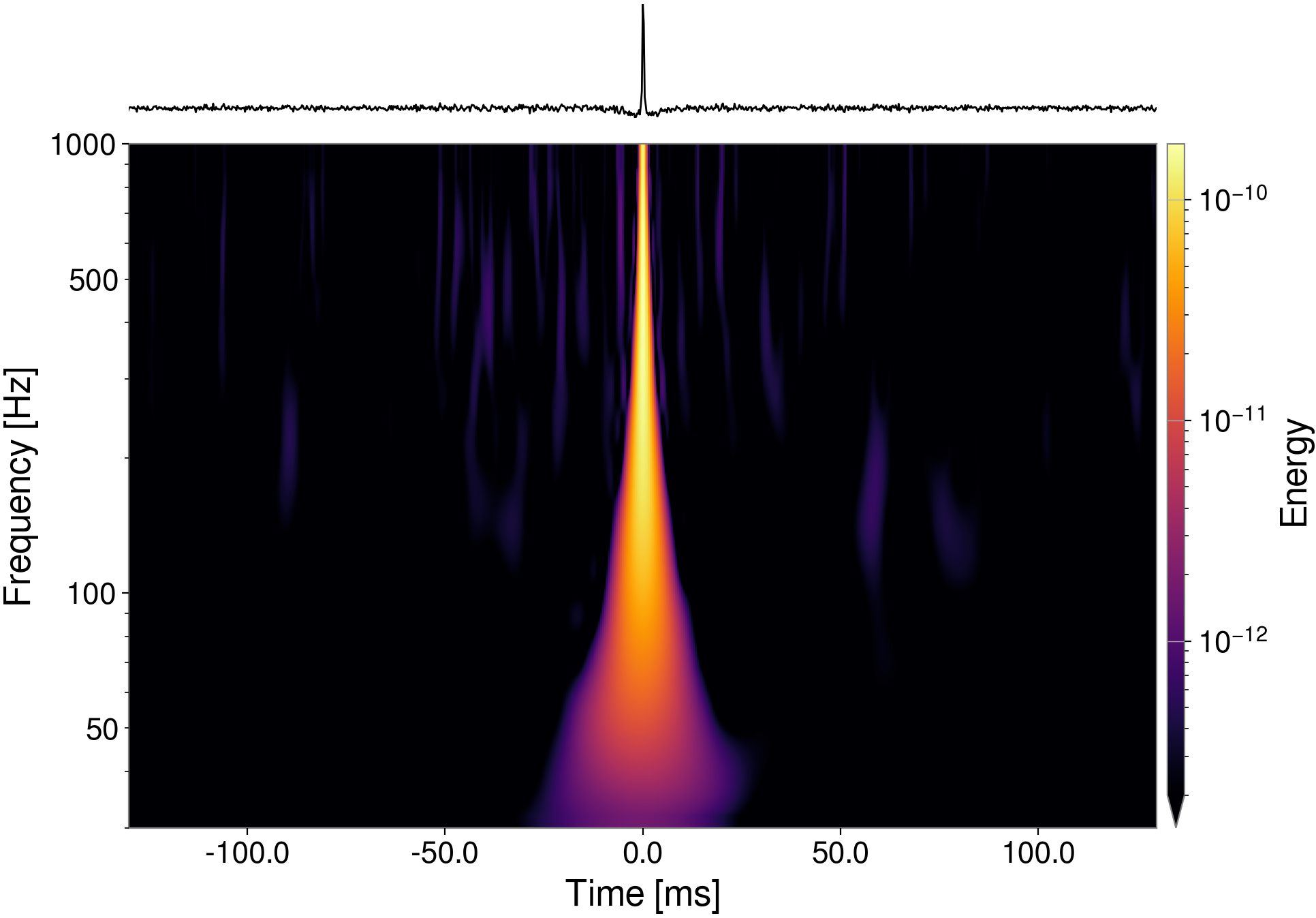}
        \label{fig:glitch_classes:blip}
    \end{subfigure}\hfill
    \begin{subfigure}[b]{0.48\textwidth}
        \caption{Koi Fish}
        \includegraphics[width=\linewidth]{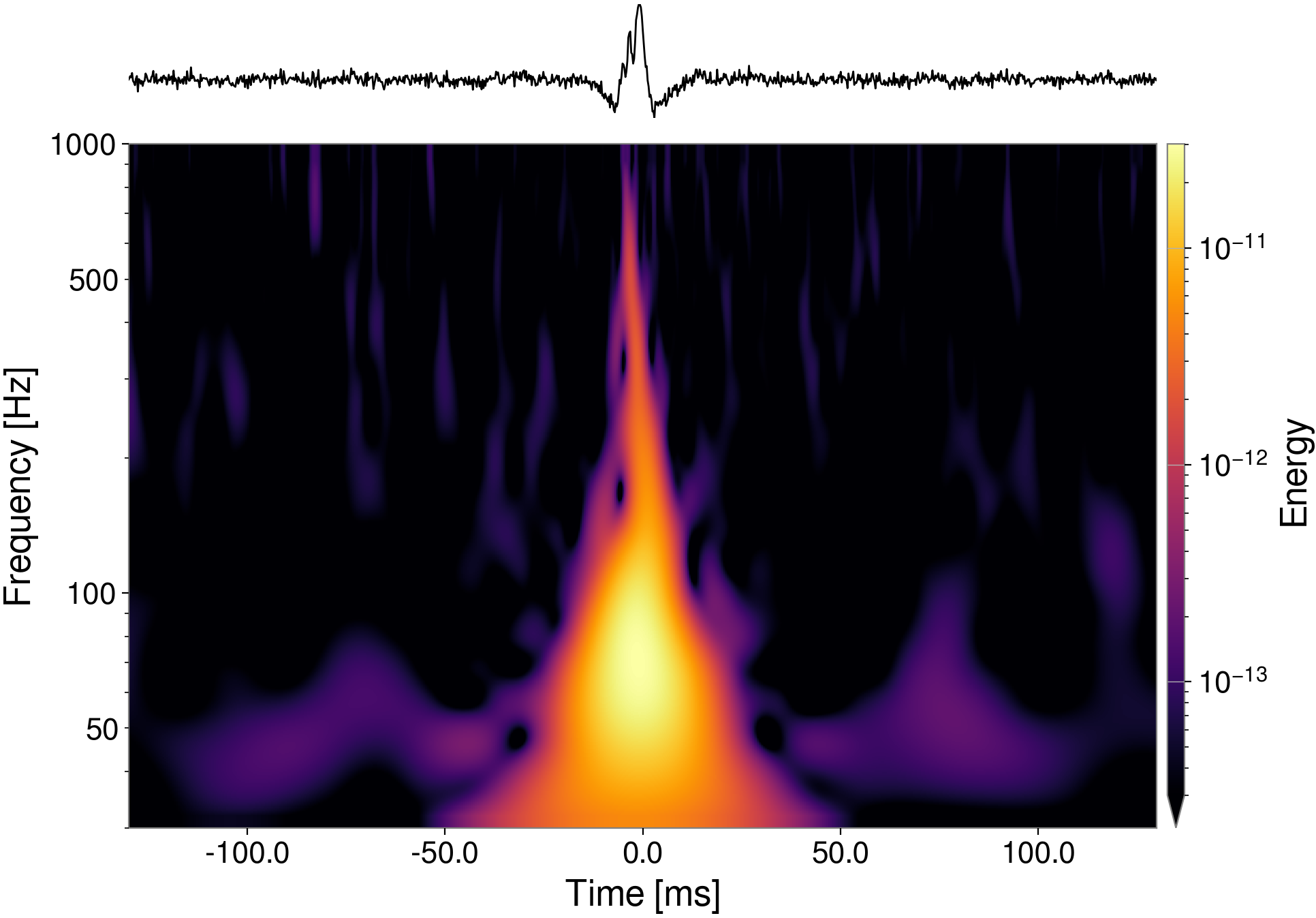}
        \label{fig:glitch_classes:koi_fish}
    \end{subfigure}

    \vspace{-1em}

    \begin{subfigure}[b]{0.48\textwidth}
        \caption{Tomte}
        \includegraphics[width=\linewidth]{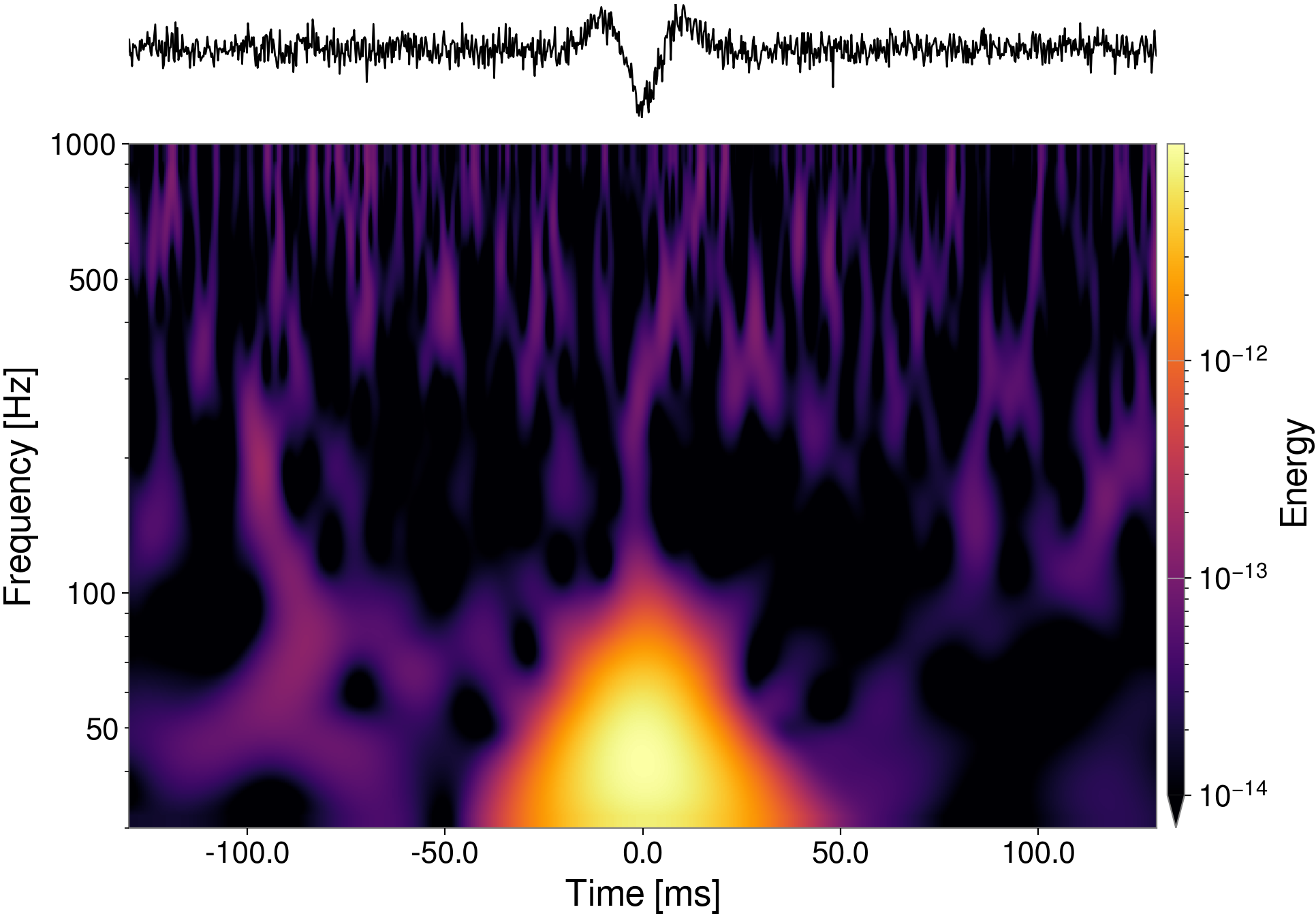}
        \label{fig:glitch_classes:tomte}
    \end{subfigure}
    \hfill
    \begin{subfigure}[b]{0.48\textwidth}
        \caption{Whistle}
        \includegraphics[width=\linewidth]{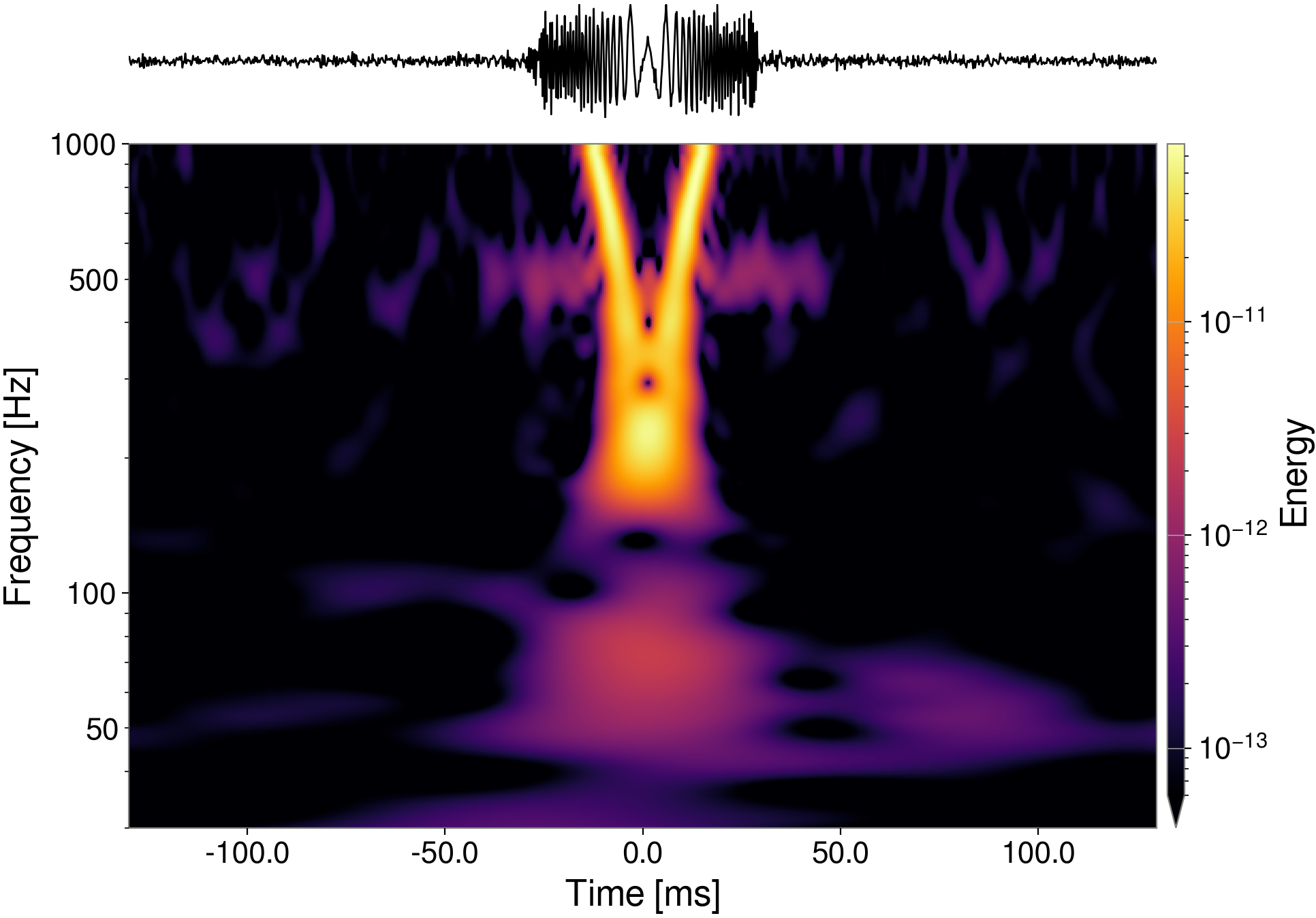}
        \label{fig:glitch_classes:whistle}
    \end{subfigure}    

    \caption{Time-frequency representations of the four families of glitches chosen for the classification demonstration: Blip (\ref{sub@fig:glitch_classes:blip}), Koi Fish (\ref{sub@fig:glitch_classes:koi_fish}), Tomte (\ref{sub@fig:glitch_classes:tomte}), and Whistle (\ref{sub@fig:glitch_classes:whistle}). A normalised time-series overlay (black) is shown above each panel.}
    \label{fig:glitch_classes}
\end{figure*}

Let us go back to the match-optimised example to elaborate on the observed behaviour of the basic reconstruction method, and to exemplify possible further improvements due to the model's interpretability. As explained in Section~\ref{sec:simple-reconstruction}, the overlapping windows into which the input strain is split and denoised retain their original $\ell_2$ norm. This is the reason why the denoised regions with the highest energy density coincide with the original input strain.
While this method retains the evolution of the observed signal's energy over time with a simple and efficient algorithm, it inevitably comes with downsides. The main one is that non-astrophysical stochastic features with high-enough local spectral energy density will eventually overcome the effective discrimination threshold of the dictionary. Their reconstructions, however, remain sparse and therefore bound to display morphological traits akin to the training samples even when the underlying stochastic sources do not exhibit them.
This effect can be observed in two regions of Figure~\ref{fig:gw170817_denoisings:match_optimised_rec}: short and consecutive sweep curves from 200~Hz to $\sim$420~Hz arising from the beginning of the plot until time $-300$~ms, and several horizontal lines between 100~Hz and $\sim$250~Hz after the merger.
Such artefacts can nevertheless be mitigated and even filtered out taking advantage of the interpretability of the SDL model. It is easy to track which atoms are the main contributors to each part of the denoised signal, and in our example we observe a (rather evident) correlation between frequency region and the main parts of a CBC wave in the \textsc{PyCBC} dictionary.
This dictionary recovers high-frequency features using mostly late-inspiral and merger parts, whereas lower-frequency features come from earlier inspiral samples.
Provided that we are certain our event comes from a CBC and, for example, no other astrophysical events are present in our data, then we can use this semantic information in a second-pass step. In the case of the \textsc{PyCBC} dictionary we can weight down the contribution of high-frequency atoms (trained with merger regions) during the inspiral, as well as remove the contribution of low-frequency atoms (trained with early-inspiral regions) after the merger. In the case of the CoRe dictionary, because it is trained with larger atoms and shorter signals comprising mostly merger regions, its reconstruction can be limited to the known merger region, although fine-tuning the contribution of its atoms is not as straightforward. Nevertheless, the dictionary proved to be particularly discriminant, especially with lower $\lambda_\text{den}$ values, only providing a non-zero reconstruction where the original merger was located. Hence, it is viable to use the CoRe dictionary to infer the position of the merger before using the \textsc{PyCBC} dictionary for the second pass.

\subsection{Classification of glitches
\label{sec:illustrative-examples-classification}}

To illustrate the classification method currently implemented in \textsc{clawdia}, we target four common families of instrumental glitches from LIGO's third observing run: {Blip}, {Koi Fish}, {Tomte}, and {Whistle}. Glitch times and metadata are taken from~\cite{Glanzer:2021} while strain data are retrieved from GWOSC, as in the denoising example of Section~\ref{sec:illustrative-examples-denoising}.

Since glitches lack a notion of ``clean'' waveforms analogous to templateable GWs, we bias the training set towards confidently labelled, reasonably loud examples. Concretely, we require catalogue confidence $\geq 0.95$, restrict the SNR to $[20,60]$ to avoid extreme outliers (whose morphology and spectral support can differ significantly from more typical instances), and limit durations to $<1$\,s to keep classification atoms reasonably sized. Long-lived or quasi-continuous disturbance classes (e.g., scattered light, wandering lines) are excluded. To limit class imbalance after filtering, we cap each family at 200 examples. The final dataset comprises 200 Blips, 94 Koi Fish glitches, 38 Tomtes, and 200 Whistles; representative time–frequency maps are shown in Figure~\ref{fig:glitch_classes}.

Each example is extracted as a 1\,s strain segment centred on the glitch, highpass filtered at 30\,Hz, and whitened using an ASD estimated from the preceding 100\,s of data. We deliberately omit a denoising stage to keep the set-up minimal and to isolate the behaviour of the classifier.
The dataset is split in a stratified 70/30 ratio for training and testing, respectively. 
Classification uses the \texttt{DictionaryLRSDL} interface of Section~\ref{sec:classification-with-LRSDL}.
Because, at present, this dictionary operates on fixed-length inputs without patching, atoms must capture as much contextual information as possible while also leaving room to extract enough training patches. We therefore set the atom length to $0.5$\,s.
To reduce time-shift sensitivity without including alignment heuristics, each 1\,s training segment is decomposed into four overlapping half-second windows (Section~\ref{sec:simple-reconstruction}), exposing the dictionary to multiple intra-window placements of the same morphology  and improving robustness to timing offsets at inference.

We adopt a lightweight hyperparameter search guided by the weighted $F_1$ score.
The number of class-specific atoms per class, $a_C$, is first constrained by the available number of training patches: because this remains smaller than the feature count (i.e., the atom length), the classification dictionary is \emph{undercomplete}. After accounting for the four-way time-shift augmentation (which increases patch availability), we decided to use $a_C = 100$ atoms per class.
The sparsity and Fisher terms are tuned jointly: we scan a bidimensional logarithmic grid for $(\lambda_1,\lambda_2)$ with both parameters ranging from $10^{-4}$ to $1$ and select $\lambda_1=\lambda_2=10^{-3}$ at the grid optimum.
This optimisation is performed with no shared dictionary both to simplify the optimisation process and because the training pool is limited: excessive sharing risks absorbing class-discriminative structure. Nevertheless, to quantify any benefit we perform a secondary search over $a_0$ using 10 equally spaced values in $[10,100]$ atoms and obtain a modest improvement at $a_0=30$.
The nuclear-norm weight $\eta$ is known from prior work to be comparatively insensitive over several orders of magnitude for this setting; we therefore fix $\eta=10^{-4}$ during the $(\lambda_1,\lambda_2,a_0)$ search and later confirm empirically that $\eta=10^{-2}$ and $\eta=10^{-6}$ yield indistinguishable results.

\begin{figure}
    \centering
    \includegraphics[width=0.8\linewidth]{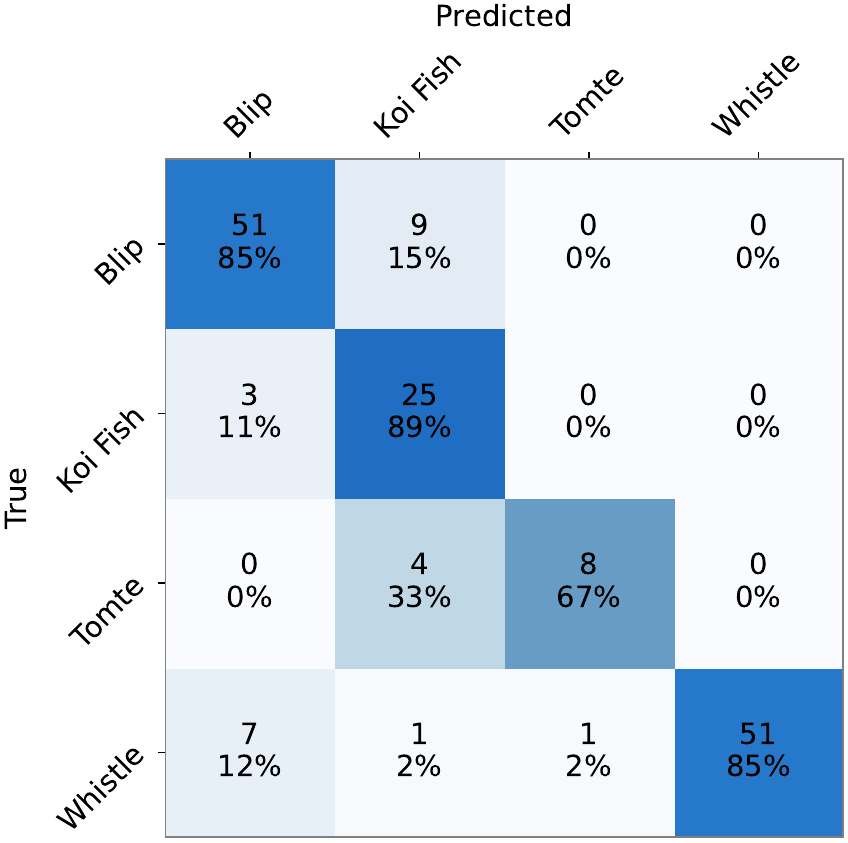}
    \caption{Confusion matrix showing the classification results over the test samples. Rows indicate the actual morphology, and columns the predicted one. Each cell indicates the absolute count of glitches and the percentage that represents over the total number of true values. To facilitates the visualisation, each cell is filled with a blue background following a linear colour scale aligned to the percentage values.}
    \label{fig:glitch_cmap}
\end{figure}

On the test carried out the classifier attains a weighted $\text{F}_1 \simeq 0.85$, with per-class behaviour summarised in the confusion matrix of Figure~\ref{fig:glitch_cmap}. 
While the low number of test samples limits the extent to which we can reach significant conclusions on the distribution of predictions, we can still comment on the overall behaviour.
Precission is relatively uniform across classes, even for the Tomte class since it only included 12 test samples.
Blips are occasionally confused with short-lived Koi Fish due to their shared compact, high-frequency content.
Tomtes are sometimes identified as longer Koi Fish, consistent with their overlapping low-frequency structure and the scarcity of Tomte training data.
Whistles are primarly confused with Blips. Although true Whistles display a very characteristic frequency drift, their mid-to-high-frequency energy can be partially synthesised by combining several time-shifted, blip-like atoms (an expectable artifact given our simple time-shift augmentation and the absence of a denoising front end). Visual inspection of misclassified cases confirms that these tend to be marginal outliers for their nominal class (e.g., unusually short or unusually broadband instances), for which a competing class dictionary (in our case, Blips) provides a slightly better match.

We consider these results encouraging given the intentionally simple configuration, the highly imbalanced training pool, and the lack of denoising.
Several straightforward refinements are expected to yield significant gains.
Introducing a denoising stage optimised to preserve class-discriminative morphology would reduce background variability prior to classification.
Decreasing the window step during training would increase the effective sample size (and the density of time-shift coverage), enabling a larger number of class-specific atoms and overcomplete class dictionaries, provided enough computing resources.
Looking ahead, the forthcoming addition of patch-based inference to the \textsc{LRSDL} dictionary will remove the fixed-length constraint and reduce atom lengths to scales better suited for the target morphologies, as discussed in Sections~\ref{sec:classification-with-LRSDL} and~\ref{sec:future-directions}.

\section{SUMMARY AND OUTLOOK\label{sec:future-directions}}

We have presented \textsc{clawdia}, a modular, open-source pipeline developed to facilitate the analysis of GW data, with a particular focus on signal denoising and classification tasks. \textsc{clawdia} is intended as a community-driven, interoperable library extensible to additional tasks, including detection and parameter estimation. This tool, implemented as a Python package, integrates SDL techniques and is designed to offer flexibility in addressing specific challenges in the field. The modularity of \textsc{clawdia} allows it to be used both as a complete pipeline for comprehensive workflows and as a library of independent functions, providing versatile options to adapt to specific processing needs and enabling future extensions or integrations with other tools. System parameters, such as segment size, sliding window step, and dictionary configurations, are adjustable, making it easy to optimise performance according to the characteristics of the dataset or the limitations of the computational environment. In order to provide an integrated workflow, \textsc{clawdia} maintains a close connection with \textsc{gwadama}, a complementary package for data preparation, signal normalisation and whitening, as well as the generation of training and testing sets from simulated signals and injected noise. 

We have demonstrated \textsc{clawdia}'s performance by denoising the signal from BNS merger event GW170817 and classifying families of instrumental glitches from LIGO's third observing run, highlighting robustness in low SNR conditions.  Further applications of the \textsc{clawdia} pipeline have demonstrated promising results in studies of supernova explosions~\cite{Powell:2024} and on the reconstruction of post-merger signals in BNS mergers~\cite{Llorens-Monteagudo:2025:bns}.

Several extensions to \textsc{clawdia} are under consideration, motivated by practical insights from its current applications and by recent developments in sparse modelling and signal processing. With respect to classification, a near-term priority is the implementation of more advanced rejection criteria for foreign signals that deviate from the modelled classes. Ongoing work is exploring the use of generic dictionaries trained for CBC classification or anomaly detection. These dictionaries would operate in the early stages of the classification pipeline, improving robustness in open-set conditions. Related to this effort is the exploration of confidence-based rejection strategies, including methods derived from conformal prediction to quantify prediction uncertainty~\cite{Malz:2025,Ashton:2024,Ashton:2025}.

In parallel, the \textsc{LRSDL} model will be extended to support arbitrary-length signals via patch-based training. This would lift the current burden of dealing with potentially long atoms and, therefore, extremely limited training data, as well as the intrinsic problem of signal alignment. The central challenge, however, is aggregating predictions across windows during inference. Several strategies are under evaluation, such as confidence-weighted voting schemes~\cite{Kuncheva:2014,Dogan:2019} and sparsity-informed attention models~\cite{Vaswani:2023,Iwanaga:2025,Liu:2023}, aiming to assign class predictions based on the most informative segments of each signal.

Beyond classification, improvements to sparse representations are also being studied. One direction is the use of frequency-targeted dictionaries to enhance spectral coverage, particularly in applications involving broadband signal classes with distinct substructure across the spectrum. This approach consists on either using multiple specialised dictionaries, each tuned to different spectral bands, or fine-tuning the atoms within a single dictionary to promote a custom frequency-weigthed population. It could be particularly advantageous for signals with features distributed across widely separated frequency bands or varying spectral content across samples.

Another important goal is adapting the framework to varying noise conditions and multi-detector configurations. Adaptive or noise-aware dictionary learning methods, capable of automatically adjusting to different SNR levels or detector responses, are being considered to improve performance without requiring separate model configurations. Similarly, multi-detector dictionaries are being explored for coincident signals across multiple interferometers. The main idea involves encoding synchronised strain data as multichannel arrays, adapting the dictionary structure analogously to colour-channel representations in image dictionaries. 

Physically informed dictionary learning (PIDL) represents a further potential direction. Recent work has demonstrated the incorporation of physical constraints into the dictionary learning process through customised regularisation terms, enforcing atoms to encode features consistent with underlying physical models~\cite{Tetali:2019,Liu:2024,Damiano:2025}. While direct application to highly nonlinear or stochastic GW signals may be limited in the time domain, adaptations to frequency or image-based representations could make this approach feasible.

The scope of the framework may also be extended towards additional GW signal types, including continuous waves, stochastic backgrounds, and highly eccentric compact binaries. Support for overlapping signals such as glitch--GW and GW--GW superposition is another prospective application. This would require advanced methods for signal separation or multi-source reconstruction, particularly in the context of burst-like or post-merger signals.

Complementary to these efforts, future development includes the integration of alternative data representations. While time-domain signals are currently the only format supported in \textsc{clawdia}, other representations---such as amplitude--phase decomposition or Fourier-domain transforms---may offer advantages for signals with complex or stochastic morphology. By distributing information across multiple channels, these formats could facilitate the extraction of meaningful features by the dictionary.

From a methodological standpoint, optimisation strategies and workflows may be implemented within \textsc{clawdia}, as an attempt to standardise these procedures as much as possible, with proper documentation. Besides implementing our current approach (hinted at by the guidelines in Section~\ref{sec:clawdia-optimisation-guidelines}, and used in our last study~\cite{Llorens-Monteagudo:2025:bns}), we are exploring new methods that may be incorporated to the framework in the future. Such is the case of curriculum-based training strategies~\cite{Bengio:2009,Marx:2025}, which are also being explored. These involve progressive learning schemes in which dictionaries are first trained on clean signals with low regularisation, and subsequently refined on noise-injected signals under stronger sparsity constraints. Such staged learning procedures could improve generalisation, especially in low-SNR regimes. Finally, optimisation of \textsc{clawdia}'s computational efficiency through vectorised operations, modular parallelism, and compiled extensions is expected to be progressively implemented. This will become necessary for more ambitious applications, to improve scalability to large datasets and real-time analysis.

\begin{acknowledgments}
This work is supported by the Spanish Agencia Estatal de Investigaci\'on (grants PID2021-125485NB-C21 and PID2024-159689NB-C21) funded by MCIN/AEI/10.13039/501100011033 and ERDF A way of making Europe. Further support is provided by the Generalitat Valenciana (CIPROM/2022/49) and  by  the  European Horizon  Europe  staff  exchange  (SE)  programme HORIZON-MSCA-2021-SE-01 (NewFunFiCO-101086251).

This research has made use of data or software obtained from the Gravitational Wave Open Science Center (gwosc.org), a service of the LIGO Scientific Collaboration, the Virgo Collaboration, and KAGRA. This material is based upon work supported by NSF's LIGO Laboratory which is a major facility fully funded by the National Science Foundation, as well as the Science and Technology Facilities Council (STFC) of the United Kingdom, the Max-Planck-Society (MPS), and the State of Niedersachsen/Germany for support of the construction of Advanced LIGO and construction and operation of the GEO600 detector. Additional support for Advanced LIGO was provided by the Australian Research Council. Virgo is funded, through the European Gravitational Observatory (EGO), by the French Centre National de Recherche Scientifique (CNRS), the Italian Istituto Nazionale di Fisica Nucleare (INFN) and the Dutch Nikhef, with contributions by institutions from Belgium, Germany, Greece, Hungary, Ireland, Japan, Monaco, Poland, Portugal, Spain. KAGRA is supported by Ministry of Education, Culture, Sports, Science and Technology (MEXT), Japan Society for the Promotion of Science (JSPS) in Japan; National Research Foundation (NRF) and Ministry of Science and ICT (MSIT) in Korea; Academia Sinica (AS) and National Science and Technology Council (NSTC) in Taiwan.
\end{acknowledgments}

\appendix

\section{GWADAMA\label{sec:appendix-gwadama}}

\textsc{gwadama} ({\bf G}ravitational-{\bf Wa}ve {\bf Da}taset {\bf Ma}nager) is a Python toolbox developed alongside \textsc{clawdia} to accelerate the management and preparation of datasets for GW data analysis. Originally an internal utility of \textsc{clawdia}, it has evolved into a standalone library, providing a transparent abstraction layer between raw data and user-facing code. Its primary goal is to automate common operations such as preprocessing, injection, metadata handling, and structured data access across a growing variety of use cases and scenarios.

Internally, signals are represented as individual one-dimensional \textsc{NumPy} arrays, stored in nested Python dictionaries alongside their metadata and labels. This avoids zero-padding variable-length strains or retaining long stretches of uninformative background solely to populate a rectangular array, reducing peak memory usage and unnecessary data movement. When a batch format is required, \textsc{gwadama} converts on demand to fixed-shape arrays (padding only when needed) so operations like frequency filtering and whitening can run per signal or be grouped into padded mini-batches when multithreading is advantageous.

The core of \textsc{gwadama} is built around a small family of dataset classes designed for specific scenarios. The \texttt{Base} and \texttt{BaseInjected} classes provide a common interface for clean and noise-injected datasets, respectively, and act as the generic entry points. These are extended by \texttt{SyntheticWaves} and \texttt{InjectedSyntheticWaves}, which generate and manage artificial datasets (currently sine-Gaussians, Gaussians, and ring-downs), and by \texttt{CoReWaves} and \texttt{InjectedCoReWaves}, which handle waveforms from the CoRe numerical-relativity database of BNS merger simulations~\cite{CoRe:2023}, including download, resampling, frame projection, merger alignment, and SNR-controlled injection. New dataset classes are expected to emerge organically as new signal types and workflows are incorporated.

To support data preparation, \textsc{gwadama} provides a collection of signal-processing tools across several dedicated modules. The \texttt{tat} (Time Analysis Tools) module includes utilities for resampling, generating and padding consistent time arrays, several window and filter functions, and custom convolution and whitening routines adapted from \textsc{GWpy} in simplified form to avoid adding it as a dependency. The \texttt{fat} (Frequency Analysis Tools) module implements high-pass and band-pass filters, instantaneous-frequency estimation with optional phase corrections, and signal-to-noise calculations. The \texttt{synthetic} module provides Gaussian-noise generation (including non-white noise via PSD/ASD profiles) and injection utilities, together with waveform generators for common burst-like morphologies. These tools are integrated within the dataset classes to facilitate reproducible workflows with minimal user intervention.

\paragraph*{Whitening (implementation note).}
Whitening is a conditioning step used throughout this and prior SDL work with \textsc{gwadama}. Because it reweights frequency content and thus the sparsity and atom selection, small implementation choices can affect results. For reproducibility we state the defaults used in this paper: unless an ASD is provided, one is estimated \emph{per strain} via Welch with median averaging, using a Hann window and segment length equal to the FIR length. Optional normalisation during whitening is available through \texttt{normed} with modes \texttt{peak}, \texttt{l2}, and \texttt{mad}. For injected datasets, the frequency cut-off parameter also acts as the high-pass applied to clean signals prior to injection, keeping the assumed PSD/ASD consistent.

Future plans include the standardisation of interfaces, improved support for multi-detector data, and the development of tools to simulate real-time analysis scenarios. The codebase is publicly available on GitHub\footnote{
    \url{https://github.com/MiquelLluis/GWADAMA}
} and includes API documentation styled after \textsc{NumPy}'s.

\bibliographystyle{apsrev4-2}
\bibliography{references}

\end{document}